\pdfoutput=1
        \documentclass[11pt,preprint,nocopyrightspace,onecolumn]{sigplanconf}%\documentclass[9pt,preprint,nocopyrightspace]{sigplanconf}
        \conferenceinfo{EuroSys '15}{April 21--24, 2015, Bordeaux, Bordeaux, France}
        \copyrightyear{20yy} 
        \copyrightdata{978-1-nnnn-nnnn-n/yy/mm} 
        \doi{nnnnnnn.nnnnnnn}
        
\def\paperversion{tr}
\newif\ifsinglecolumn\singlecolumnfalse
\newif\ifwidemargins\widemarginsfalse
\newif\ifwarning\warningfalse
\newif\ifshowcomments\showcommentsfalse
\newif\ifblinded\blindedfalse
\newif\ifreport\reportfalse
\newif\ifcopyrightspace\copyrightspacefalse
\newif\ifacknowledgments\acknowledgmentsfalse
\newif\ifshowpagenumbers\showpagenumberstrue
\newif\iffinalformat\finalformatfalse

\IfFileExists{.blinded}{\blindedtrue}

\ifx\paperversion\xxxxundefined
\PackageError{paperversions}{*** No valid document version was specified.
Macro paperversions must be be defined as one of (markup, draft,
local, submission, final, web, tr, trdraft)}
\fi

\def\xxversion{\csname xx\paperversion\endcsname}
\newif\ifsawversion\sawversionfalse

\ifcase\xxversion\relax
    \widemarginstrue
    \singlecolumntrue
    \warningtrue
    \sawversiontrue
\or %draft
    \warningtrue
    \showcommentstrue
    \sawversiontrue
\or %local
    \warningtrue
    \showcommentsfalse
    \blindedfalse
    \sawversiontrue
    \acknowledgmentstrue
\or %submission
    \sawversiontrue
\or %final
    \blindedfalse
    \sawversiontrue
    \copyrightspacetrue
    \acknowledgmentstrue
    \showpagenumbersfalse
    \finalformattrue
\or %tr
    \singlecolumntrue
    \blindedfalse
    \sawversiontrue
    \reporttrue
    \acknowledgmentstrue
\or %trdraft
    \blindedfalse
    \singlecolumntrue
    \showcommentstrue
    \sawversiontrue
    \reporttrue
\or %finaldraft
    \blindedfalse
    \showcommentstrue
    \copyrightspacetrue
    \acknowledgmentstrue
    \blindedfalse
    \sawversiontrue
    \finalformattrue
\or %web
    \blindedfalse
    \sawversiontrue
    \copyrightspacetrue
    \acknowledgmentstrue
    \finalformattrue
\fi

\ifsawversion
\else
\fi

\let\xxversion=\undefined

\usepackage{times-lite,txtt,graphicx,color}
\usepackage{amsmath, amssymb, stmaryrd}
\usepackage{ttquot,utf8}
\usepackage{appendix}
\usepackage{arydshln}
\ifreport\relax
    
\else
    \usepackage{tightheaders}
\fi
\ifshowcomments\relax\else
    \PassOptionsToPackage{disabled}{authcomments}
\fi
\usepackage[checkNoComments]{aplcomments}
    \newcommenter{RVR}{0.2,0.5,1}
\usepackage{subfigure}
\usepackage{hyperref}
\usepackage{paralist}
\usepackage{mathpartir}
\hypersetup{
	colorlinks,
	urlcolor=black,
	citecolor=black,
	filecolor=black,
	linkcolor=black
}
\usepackage{color}
\usepackage{listings}
\usepackage{harpoon}
\usepackage{xspace}
\showpagenumberstrue
\ifshowpagenumbers
    \def\thepage{\arabic{page}}
\fi

\renewcommand{\floatpagefraction}{0.75}

\renewcommand\paragraphmark[1]{.}

\ifreport
\title{Distributed Protocols and Heterogeneous Trust: Technical Report}
\else
\title{Distributed Protocols and Heterogeneous Trust}
\fi

\newcommand\draftwarning{
\ifwarning \vspace{1ex} {\it\color{red}{\large (Draft---please do not distribute)}} \else \fi
}

\ifreport
    \authorinfo{Isaac C. Sheff}{Cornell University}{isheff@cs.cornell.edu}
    \authorinfo{Robbert van Renesse}{Cornell University}{rvr@cs.cornell.edu}
    \authorinfo{Andrew C. Myers}{Cornell University}{andru@cs.cornell.edu}
\else
    \ifblinded
    \authorinfo{EuroSys 2015 \#127, 14 pages (12 of content, 2 of references) \draftwarning}{}{}
    \else
      \authorinfo{Isaac C. Sheff}{Cornell University}{isheff@cs.cornell.edu}
      \authorinfo{Robbert van Renesse}{Cornell University}{rvr@cs.cornell.edu}
      \authorinfo{Andrew C. Myers}{Cornell University}{andru@cs.cornell.edu}
    \fi
\fi

\ifblinded
\else
\fi
\usepackage{makeidx}  % allows for indexgeneration
\newcommand{\p}[1]{{\left({{#1}}\right)}}
\newcommand{\cb}[1]{{\left\{{{#1}}\right\}}}

\newcommand{\abs}[1]{{\left|{{#1}}\right|}}

\newcommand{\tb}[1]{{\textrm{\textbf{{#1}}}}}
\newcommand{\join}[0]{{\sqcup}}
\newcommand{\meet}[0]{{\sqcap}}
\newcommand{\less}[0]{{\sqsubseteq}}
\newcommand{\actsfor}[0]{{\succeq}}
\newcommand{\avail}[0]{{\xleftarrow{A}}}
\newcommand{\integ}[0]{{\xleftarrow{I}}}
\newcommand{\noinbf}[1]{{\smallskip\noindent\textbf{{#1}}}}

\usepackage{sidecap}% for figures with captions on the side.
\usepackage[boxed,linesnumbered]{algorithm2e}% for pseudocode
\usepackage{tikz} % for lattice diagram
\usepackage{mathtools}

\newcommand{\notless}[0]{{\mathrlap{\, /}\less}}

\usepackage{overrides}
\begin{document}
\maketitle
\begin{abstract}
The robustness of distributed systems is usually phrased in
terms of the number of failures of certain types that they
can withstand. However, these failure models are too crude to
describe the different kinds of trust and expectations of participants
in the modern world of complex, integrated systems extending across
different owners, networks, and administrative domains.  Modern
systems often exist in an environment of heterogeneous trust, in 
which different participants may have different opinions about the
trustworthiness of other nodes, and a single participant may consider
other nodes to differ in their trustworthiness.  We explore
how to construct distributed protocols that meet the requirements of all participants,
even in heterogeneous trust environments. The key to our approach
is using lattice-based information flow to analyze and prove
protocol properties.  To demonstrate this approach,
we show how two earlier distributed algorithms can be generalized to
work in the presence of heterogeneous trust: first, Heterogeneous Fast
Consensus, an adaptation of the earlier Bosco Fast Consensus
protocol; and second, Nysiad, an algorithm for converting
crash-tolerant protocols to be Byzantine-tolerant.  Through
simulations, we show that customizing a protocol
to a heterogeneous
trust configuration yields performance improvements over the
conventional protocol designed for homogeneous trust.

\end{abstract}

\setcounter{page}{1}

%% \mainmatter % tells llncs to use real page numbers starting here.
\section{Introduction}
\newcommand{\crash}{crash}

\label{introduction}
\label{example-intro}
Fault tolerance
%\ACM{was: failure tolerance, but this seems like the more standard term}
%\ICS{Sounds good.}
is critical for distributed systems.
Traditionally, distributed systems and protocols are
designed around the ability to tolerate some number of failures,
sometimes differentiated by type, such as crash or Byzantine
\cite{Schneider83a,Schneider90,byzantinegenerals,osdi99}.
For well-studied problems such as consensus, lower bounds are 
traditionally expressed in terms of the number of participants needed to
tolerate some number of failures $f$
\cite{byzantinegenerals,Bracha85,Brasileiro2001,Song2008}.
In our increasingly interconnected world, however, systems must routinely 
operate across locations and between different owners, requiring a richer 
notion of what failures are possible. In complex systems integrated
across administrative domains (that is, federated systems), different
participants in the system may not even agree on 
what types of failures may occur or where in the system~\cite{survivor-sets,foley1,integrity,Steiner1993}.

\begin{figure}
\centering
\ifreport
  \includegraphics[width=\textwidth]{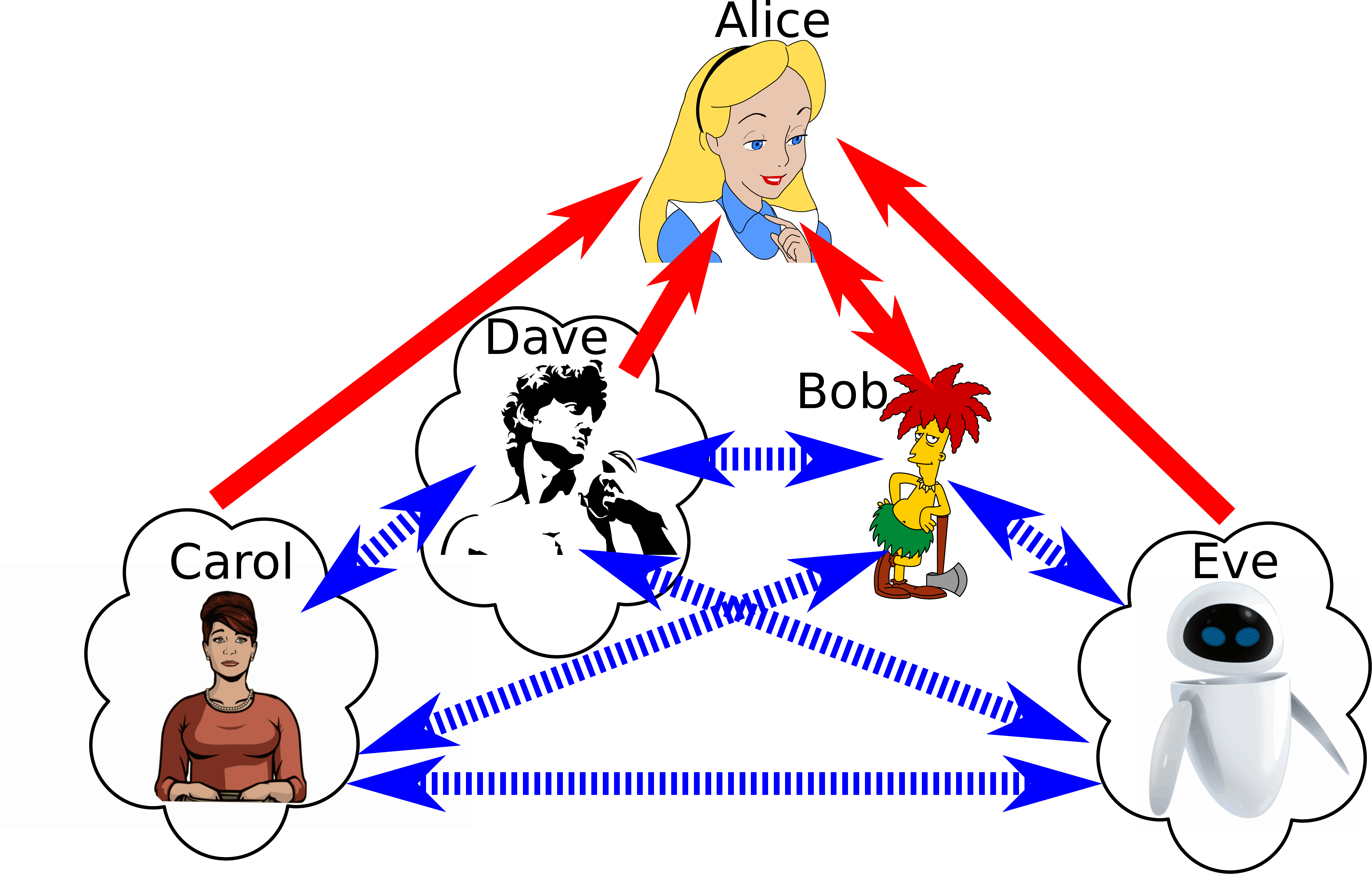}
\else
  \includegraphics[height=50mm]{characterscrasharrow.pdf}
\fi
\caption{
Red solid arrows represent participants' belief that another participant might 
lie (integrity failure), while blue striped arrows represent participants' 
belief that another participant may crash (availability failure).
Everyone but Alice believes at most one crash may occur, and Alice may lie in
addition.
Alice believes only Bob can fail, although he may fail in any manner.
}
\label{fig:example}
\end{figure}

For example, suppose Alice is building a new app, a competitor to
Bob's.  However, due to the nature of the application, it is best
if Alice and Bob's apps agree (quickly) on some things.  Alice's app
works with servers maintained by Carol, Dave, and Eve.
Due to strong records and reliably enforced SLAs, Alice is willing to
allow her program to fail if Carol, Dave, or Eve do. She does not
tolerate their failure.  She fears, however, that Bob may lie to her.
On the other hand, Bob, Carol, Dave, and Eve do not know Alice (her
app is new, after all), and believe she may lie or crash.  As
established businesses with contracts and track records, however, they
believe each other to be honest, although they have a healthy
tolerance for at most one crash among themselves.

This trust configuration, depicted in Figure~\ref{fig:example}, is
much more complex than in the traditional
models where participants uniformly agree on the maximum number of
crash failures and the maximum number of Byzantine failures. In this
paper, we explore the possibility of more general distributed
protocols that take into account and even exploit such complex, heterogeneous trust.
%\ACM{protocols or algorithms?}
%\ICS{I suppose we use ``algorithms'' in the abstract. In reality, either works. I just think ``algorithm'' sounds fancier. When I hear ``protocol,'' I think IP, not Paxos.}
% \ACM{Still confused about algorithm vs protocol. 'Fancier' is not
% better, ceteris paribus. But the literature does talk about the 
% Bosco 'algorithm' and the Nysiad 'algorithm'... Robbert, can you
% decide about this?}

\subsection{Contribution}

Heterogeneous trust presents a challenge for designing fault-tolerant protocols.
Our key idea is to use information flow methods to reason
about the integrity and availability properties of distributed
systems. Prior work on information flow methods has mostly addressed
confidentiality properties of systems~\cite{sm-jsac}, though some
prior work~\cite{zznm02,zcmz03,zm05,zm14} has addressed integrity and
availability properties in a limited way. However, our work exploits
information flow in a more general and sophisticated way.

We construct and analyze fault tolerant protocols by tracking the
integrity and availability of flows of information through the
protocol.  Intuitively, each participant characterizes its assumptions
about the availability and integrity of system components as a
_label_ drawn from a rich lattice that expresses the trust of
participants.  The labels are
expressive enough to represent various combinations of crash and/or
Byzantine failures, as expected by the various participants.
Using these labels, it is then possible to analyze under what conditions 
the results of running a protocol have an availability and integrity acceptable
to each participant. 
%This label can also be seen as representing the power of
%an ``attacker'' from the viewpoint of the participant. 

When participants have different opinions about the trustworthiness of
the system components, a new phenomenon arises: some participants'
trust assumptions may be violated while others' assumptions still hold. We
consider a protocol to be correct only if
the violation of one participant's trust assumptions
cannot damage the availability or integrity of the system
as viewed by any participant whose assumptions have not been violated.
Naturally, not all protocols can be run with all possible
configurations of trust among participants;
each protocol has minimal requirements concerning those
participants' trust in each other.

As an example of this style of synthesis,
we present Heterogeneous Fast 
Consensus, a generalization of the ``Bosco'' protocol
\cite{Song2008}. 
This generalization 
achieves the same bounds when traditional homogeneous
trust assumptions are used, but the protocol is capable
of operating with heterogeneous assumptions.
We generalize the traditional properties of a consensus
protocol—Unanimity, Validity, Progress, and Termination—for heterogeneous trust
environments~\cite{Bracha85,Song2008}.
We explain this protocol's requirements, and prove it satisfies these
properties. Using simulations,
we also demonstrate that Heterogeneous Fast Consensus offers
significant advantages in speed and resource requirements
when compared to the Fast Consensus protocol it generalizes.

As a second example, we develop
a generalization of Byzantine tolerant Ordered, Asynchronous, Reliable 
Broadcast (OARcast) and Nysiad, an algorithm for converting
{\crash}-tolerant protocols into Byzantine-tolerant protocols~\cite{Ho2007,Ho2008}.
This generalization demonstrates that in some cases, Availability and Integrity can affect
each other in counterintuitive ways.  For example, Nysiad includes
cases in which a participant cannot affect a value's integrity, but
can make it unavailable by lying.  

\subsection{Related Work}

Others have looked into richer notions of failure, including generalizing 
$f$-failures to failure-prone and survivor sets \cite{survivor-sets}, or 
mixing availability (\textit{omission}, or \textit{Crash}) and integrity (\textit{commission}, 
or \textit{Byzantine}) failures 
\cite{Meyer1991,Garay1992,Siu1998}. 
Some work operates on expanded failure models, including notions of selfish but
not malicious
%\ACM{was 'predicable', a typo I assume. How are they predictable?} 
%predicatable in that you can predict they'll be selfish? I changed it.
participants, sometimes mixed with other kinds of 
failure~\cite{bar-sosp,abraham2006}.
A distinctive feature of our work is the removal of the assumption that
all participants share the same notion of possible failures. 
Properties of the form ``as long as the failure assumption is not violated,
it is guaranteed that \dots'' can be generalized to explain which guarantees
can be made for which participants.

This work exists at the intersection of information flow analysis, a technique 
traditionally applied through programming languages (e.g., \cite{jif}), and distributed 
systems theory.
%Distributed languages like Fabric 
%kinds of protocols may be implemented \cite{fabric09}.
Most prior research on language-based security has not concerned itself with fault
tolerance, but there are exceptions. 
Zheng has explored using information flow to reason about
availability~\cite{zm05} and integrity~\cite{zcmz03} properties,
including in distributed systems~\cite{zlt-thesis,Zheng2014}, but
the focus has been on linguistic mechanisms and simple quorum-based
protocols. 
Walker et al.~\cite{lambda-zap} design a lambda calculus formalizing
the possibility of an integrity fault on a single machine. 

Consensus, of course, is a widely-studied topic under a variety of failure 
models~\cite{jaffe2012,fastpaxos,paxos,Song2008}.
Our particular generalization of the Bosco Byzantine fast consensus protocol~\cite{Song2008}
serves as the first known
example of consensus under heterogeneous mutual distrust. 
% \ACM{Added 'heterogeneous' here because it seems to me that any
% Byzantine consensus protocol already deals with mutual distrust.}

%survivor / fail-prone sets

%various mixed-failure (omission and commission) models

%Lantian's quorums (PLAS 2014 included)

%required shout-outs to everything consensus?

\section{System model}
\label{system}
\subsection{Network}
%async
Our trust model is similar that used in other models of
distributed systems~\cite{Schneider90,Song2008,paxos}.
We assume an asynchronous network 
environment in which any participant can communicate with any other. 
There is no guarantee, however, that participants trust each other.
We assume that the network is, or can be made, reliable. 
That is to say, our participants are assumed to have whatever message resending
protocols (e.g., \cite{TCP,Afek1994}) are necessary to guarantee that the only
case in which a message never arrives is the case in which the sender or 
receiver have failed in some way. 
%compare to previous assumed models (pretty similar)
Additionally, we assume that faulty participants cannot forge the source identity
of a message sent by a correct participant; cryptographic signatures
are one way to achieve this.

% randomized / fair links
In order to guarantee probabilistic termination in Heterogeneous Fast Consensus, 
we also assume that for any set
of messages, no one of which is causally before another,
%(transitive closure of ``received by a
%participant before the other is sent by that participant''),
if the set
is sent repeatedly, the network will eventually deliver them in each possible
order.
%This is similar to Bracha and Toueg's assumption that ``we want every possible
%view to have some fixed probability of being the one seen''~\cite{Bracha1983}.
\ACM{Removed text about Bracha and Toueg's assumption, which didn't
seem important. But I could be wrong.}

\subsection{Heterogeneous Trust}
%heterogeneous trust (interesting bit, name example here)
%``availability'' for crash failures, ``integrity'' for Byzantine
Each participant has its own assumptions concerning the availability of the
system (who might crash), and the integrity of the system (who might lie, or
do something other than correctly execute the protocol). 
These assumptions can be thought of as describing what ``attacker'' is
expected by each protocol participant.
The limits of this attacker's power can be captured using information-flow labels 
(See section~\ref{ifc}). In a system with heterogneous trust,
participants that have not failed can then be characterized as either
\textbf{gurus} or \textbf{chumps}:
%chumps and gurus
\begin{trivlist}
\setlength\itemsep{2pt}
\item \textbf{Gurus}\label{gurus}
are participants who function correctly and whose trust assumptions are not violated.
By definition, no set of failures that actually occur can violate the
availability or integrity expectations of a guru. 
For most protocols, most guarantees made pertain to gurus.
\item \textbf{Chumps}\label{chumps}
are participants who function correctly---they obey the prescribed protocol and
do not crash---but whose trust assumptions have been violated.
In traditional failure-tolerant systems with homogeneous trust, all participants make the
same trust assumptions, so either all correct participants are chumps
(in which case few or no guarantees are made), or everyone is a guru.
Here, we must be more nuanced.
Unsurprisingly, chumps may receive ``wrong'' results. 
\end{trivlist}

Recall the example from section~\ref{example-intro}.
Five participants have a trust configuration in which everyone but
Alice tolerates one crash among themselves, and tolerates Byzantine
behavior from Alice. Alice tolerates failures only from Bob, but tolerates
Byzantine failures on his part.

% THIS FOOTNOTE DOES NOT APPEAR TO SHOW UP AT ALL RIGHT NOW.
%\ICS{This footnote doesn't actually show up anywhere. That's bad.}
%\footnote{Image credit:
%Alice: \url{http://inukagome134.deviantart.com/art/Alice-in-Wonderland-SVG-vector-286906042}\\
%Bob: \url{http://all-free-download.com/free-vector/vector-cartoon/sideshow\_bob\_01\_the\_simpsons\_51391.html}\\
%Carol: \url{http://oneofus.net/2014/01/archer-season-5-a-few-secrets-declassified/}\\
%Dave: \url{http://www.freepik.com/free-vector/red-black-and-white-portrait-of-david-vector-illustration\_511533.htm}\\
%Eve: \url{http://www.clker.com/clipart-cloud-outline.html}\\
%Cloud: \url{http://www.clker.com/clipart-cloud-outline.html}
%}

% probably not necessary:
%This example is illustrated in Figure~\ref{fig:example}

Suppose the participants wish to achieve consensus on data for Alice's app 
quickly, ideally in a single round in the usual case.
For example, they might want to use the Bosco Byzantine consensus
protocol, which can achieve consensus quickly. For Bosco, however,
at least nine participants would be required, since some participants believe 
at least one Byzantine failure can occur, in addition to one crash.
The five current participants would have to recruit others with relevant
trust properties. In fact,
tolerating just one Byzantine failure requires at least 6 participants for any
one-communication-step
consensus protocol tolerating $f$ failures~\cite{Song2008}.
As we show in section~\ref{heterogeneous-consensus}, it is
nevertheless possible to create a variant of the Bosco fast consensus
protocol that satisfies the requirements of all five participants,
with _no_ additional participants.

\section{Information flow policies}
\label{ifc}
\subsection{Labels}
Information flow control offers a way to reason about the properties
of information in a system.  While most prior work on information flow
has concentrated on proving confidentiality properties, it is also
possible to reason about the integrity~\cite{po95,ml-tosem} and
availability~\cite{zm05} of information. In this work,
we focus on availability and integrity of the information used in
distributed protocols, and leave confidentiality concerns to future
work.

To support the analysis of information flow, all information in the
system is assigned a label drawn from a lattice of labels that express
information security requirements for the labeled
information.~\cite{ml-tosem,cm06}. As information flows through the
system, its label (ordinarily) moves only upward in the lattice.  With
dynamic information flow control, this label is represented explicitly
at run time, whereas with static information flow control, it is
merely a compile-time aspect of the information. In this work, we use
static information flow control, because we want to design protocols
whose properties are verified before execution.

In this work, we adapt the Decentralized Label Model
(DLM)~\cite{ml-tosem} to capture the integrity and availability
requirements of information used in protocols. The DLM is designed
for systems in which principals are mutually distrusting, which is 
ideal for the design of distributed protocols.

\subsection{Principals}

Policies are expressed in terms of _principals_, which may represent machines, 
users or other entities to whom permissions may be given. 
One principal may be trusted at least
as much as another principal. If a principal $p$ is at least as
trusted as a principal $q$, we say that $p$ _acts for_ $q$,
written $p ≽ q$.\footnote{This is essentially the same idea as ``speaks
for''~\cite{labw91} in authorization logics.}
Any action that can be taken by $q$ can also be taken by $p$, meaning
that $p$ can act with the full authority of $q$.
The universally trusted principal $\top$
can act for everyone; $\bot$ is the principal with minimal authority,
for whom everyone may act. 

_Compound principals_
are a way of expressing principals representing the actions of multiple 
principals~\cite{zm05,labw91}. In particular, we use the \textit{conjunctive principal}
$p\land q$ to represent the least upper bound of the authority of $p$ and $q$
(essentially, their combined authority) and the \textit{disjunctive principal}
$p\lor q$
to represent their greatest lower bound. Therefore, for any principals $p$ and
$q$, we have $p\land q \actsfor p \actsfor p\lor q$.
Formal rules for compound $\actsfor$ can be found in Appendix~\ref{latticeappendix}.

\ACM{I got rid of the reasoning rules. They were just standard lattice
rules, so I see no reason to terrorize the reader with them.}

\if 0
The rules for reasoning about
these compound principals are summarized
in figure~\ref{fig:compoundactsfor}.

\begin{figure}[h]
\[
\begin{array}{c}
\begin{array}{c|c|c}
(p_1 \land p_2) \actsfor p_1\ &\ 
p_1 \actsfor (p_1 \lor p_2)\ &\ 
\infer{p_1 \actsfor p_2\ \ \ \ p_2 \actsfor p_3}{p_1 \actsfor p_3}
\end{array}\\
\begin{array}{c|c}
\infer{p_1 \actsfor p_3 \ \ \ \ p_2 \actsfor p_3}{(p_1 \lor p_2) \actsfor p_3}\ &\ 
\infer{p_1 \actsfor p_2 \ \ \ \ p_1 \actsfor p_3}{p_1 \actsfor (p_2 \land p_3)}
\end{array}
\end{array}
\]
\caption{
Some useful rules for $\actsfor$ on compound principals.
}
\label{fig:compoundactsfor}
\end{figure}
\fi

\subsection{Policies}

A label is a set of policies, and the label is enforced exactly when
all the policies in $\ell$ are enforced.
A policy is a statement that some principal, an owner, trusts some other 
principal to affect the labeled information.
Two kinds of policies are considered here.
A policy of the form $o \integ p$ signifies that owner principal $o$ trusts
only principal $p$ (or other principals that act for $p$)
to affect the content of this information; similarly
policy $o \avail p$ means that $o$ trusts $p$ with the availability of this 
information (that is, trusts $p$ to not make it unavailable). 

This makes intuitive sense if $p$ is a host principal, but
$p$ could also be a compound principal representing multiple hosts.
For example, a policy $o \integ p ∨ q$ means that $o$ trusts both hosts $p$ and
$q$ with the integrity of the labeled data; a failure of either
principal can destroy its integrity. Since
principal $p$ acts for $p ∨ q$, a Byzantine failure of $p$
could also compromise $p ∨ q$. Conversely, the policy
$o \integ p ∧ q$ means that $o$ believes that the integrity of
the labeled data will be compromised only if _both_ $p$ and $q$
fail.
% such as a user controlling multiple hosts. 

%\noinbf{Combined Policies}

%\section{Reasoning about heterogeneous trust}
\label{reasoning}
\subsection{Ordering labels}
\label{orderinglabelssection}
One label $\ell_2$ is at least as \textit{restrictive} as another label $\ell_1$,
written $\ell_1 \less \ell_2$, if it is always permissible to use information labeled $\ell_1$
in a situation where information with label $\ell_2$ is expected.
For this to be true, $\ell_1$ must offers integrity and availability
guarantees at least as strong as those offered by $\ell_2$.
We define the \textit{no more restrictive than} relation $\less$ on labels using the
relations $\less_I$ and $\less_A$ on integrity and availability
policies.
For $l_1$ to be no more restrictive than $l_2$ in the DLM, _every_
principal must believe that the integrity and availability
requirements expressed by $l_2$ are enforced by $l_1$. For example,
we have $\{o\integ p\} ⊑ \{o \integ p ∨ q\}$, because the
left-hand label means that $o$ believes only $p$ has affected the
data, whereas the right-hand label also permits $q$ to affect it.
Thus, the more principals have affected some information,
the more restricted future use of the information becomes.

Principals are only responsible for enforcing policies that they own,
but if $p_1 \actsfor p_2$, then $p_1$ enforces all labels that $p_2$ owns. 
If labels contain multiple policies owned by multiple different
principals, principals may have different views of the ordering on those labels.
That is why information flow is considered acceptable only when the
ordering on the labels is acceptable according to the view of _every_
principal.

The formal definition of the relationship $l_1 ⊑ l_2$ also takes into
account the trust relationships among principals~\cite{cm06}. Because
of trust relationships, the most restrictive integrity policy is $\bot \integ
\bot$, since all principals believe any principal could influence the
information, and the least restrictive is $\top \integ \top$, since
all principals believe that only $\top$ has influenced the information
(that is, it is always very trustworthy).  Availability works
much like integrity, with $\bot \avail \bot$ meaning anyone can
interfere with the labeled information's availability and
$\top \avail \top$ meaning that only $\top$
can stop the information from being available. 
A more formalized definition of $\less$ can be found in Appendix~\ref{latticeappendix}.
Figure~\ref{fig:lattice} illustrates the lattice of the integrity and availability labels.

%Figure~\ref{fig:ordering} defines an ordering on combined policies.
%Intuitively, like the orderings on policies,
%one label is ``no more restrictive than'' another if for any
%principal, the set of principals it believes can affect the data on the left is
%a subset of those it believes can affect the data on the right.
%This definition is similar to Chong's~\cite{chong-thesis}.

We define the notation $I\p \ell$ to mean the integrity policies of a label, and 
$A\p \ell$ to mean the availability policies of a label. The
relationship $l_1 ⊑ l_2$ holds exactly when the same relationship
holds separately on the availability and integrity policies of the
label: $\ell_1 \less \ell_2 \Leftrightarrow
{I\p{\ell_1}\less{I\p{\ell_2}}}\land {A\p{\ell_1}\less{A\p{\ell_2}}}$.

\subsection{Lattice operators}
\label{joinandmeet}
Since labels form a lattice, there are the usual lattice join
($\join$) and meet ($\meet$) operators. The join of two 
labels gives the strongest integrity and availability
that both kinds of information can flow to, and the meet gives the
weakest integrity and availability that is allowed to flow to both kinds
of information. Thus, $\join$ acts as a disjunction and $\meet$
acts as a conjunction.

For example, $\{o \integ p\} ⊔ \{o \integ q\} = \{o \integ p∨q\}$,
and $\{o \integ p\} \meet \{o \integ q\} = \{o \integ p∧q\}$. 
A more formalized definition can be found in Appendix~\ref{latticeappendix}.

The operations $\join$ and $\meet$ operate separately on the integrity
and availability components of labels:
\[
\begin{array}{c	|	c}
I\p{\ell_1\join\ell_2} = I\p{\ell_1}\join I\p{\ell_2}
\ \	
&\ \	
A\p{\ell_1\join\ell_2} = A\p{\ell_1}\join A\p{\ell_2}
\\
I\p{\ell_1\meet\ell_2} = I\p{\ell_1}\meet I\p{\ell_2}
\ \	
&\ \	
A\p{\ell_1\meet\ell_2} = A\p{\ell_1}\meet A\p{\ell_2}
\end{array}
\]

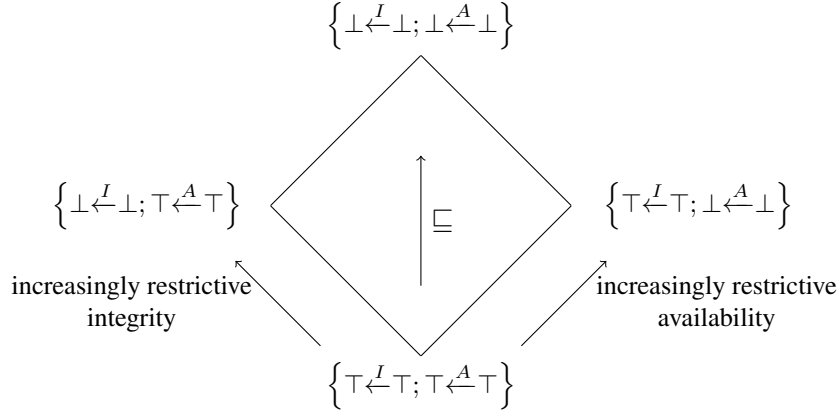
\begin{figure}[t]
\centering
\ifreport
 \begin{tikzpicture}
   [scale=.4,auto=left,every node/.style=]
\else
 \begin{tikzpicture}
   [scale=.2,auto=left,every node/.style=]
\fi
   \usepgflibrary{arrows}
   \node[scale=0.015] (d) at (11,5) {};
   \node[scale=0.015] (u) at (11,15)  {};
   \node[scale=0.015] (l) at (6,10)  {};
   \node[scale=0.015] (r) at (16,10)  {};
   \node (dd) at (11,4) {\small$\cb{\top\integ\top ; \top\avail\top}$};
   \node (uu) at (11,16)  {\small$\cb{\bot\integ\bot ; \bot\avail\bot}$};
\ifreport
   \node (ll) at (2,10)  {\small$\cb{\bot\integ\bot ; \top\avail\top}$~~};
   \node (rr) at (20,10)  {\small~~$\cb{\top\integ\top ; \bot\avail\bot}$};
\else
   \node (ll) at (0,10)  {\small$\cb{\bot\integ\bot ; \top\avail\top}$~~};
   \node (rr) at (22,10)  {\small~~$\cb{\top\integ\top ; \bot\avail\bot}$};
\fi
   \node (cd) at (11,7) {};
   \node (cu) at (11,12) {};
   \node (il) at (8,5) {};
   \node (ih) at (4.5,8.5) {};
   \node (al) at (14,5) {};
   \node (ah) at (17.5,8.5) {};
   
   \tikzstyle{every node} = [scale=1,auto=left,every node/.style=]
   
    \path (l) edge node[above] {} (u);
    \path (l) edge node[above] {} (d);
    \path (r) edge node[above] {} (u);
    \path (r) edge node[above] {} (d);
    \path[->] (cd) edge node[right] {$\less$} (cu) ;
    \path[->] (il) edge node[left] {\small \begin{tabular}{c}increasingly restrictive\\ integrity\end{tabular}\ } (ih) ;
    \path[->] (al) edge node[right] {\small \ \begin{tabular}{c}increasingly restrictive\\ availability\end{tabular}} (ah) ;
 \end{tikzpicture}
\caption{
% \ICS{The arrows for ``increasingly restrictive . . .'' don't have arrowheads in 
% Ubuntu documentviewer, but do in OSX preview and okular. I don't know why.}
The integrity and availability lattice of labels.
}
\label{fig:lattice}
\end{figure} 

\section{Reasoning about heterogeneous trust}
\subsection{Threshold synthesizers}
\label{synthesizers}
%discussion of threshold systems (citing Lantian)
Zheng and Myers introduced the concept of \textit{message synthesizers} in a 
distributed system~\cite{Zheng2014,zlt-thesis}.
A synthesizer listens for messages from a set of hosts, and based on
messages received, may produce a value with a label representing assurances
that can exceed what any one message can provide.

For example, suppose a principal $\mathtt p$ awaits the receipt of messages
$m_{\mathtt{a}}, m_{\mathtt b}, 
m_{\mathtt c}$ from principals $\mathtt a$, $\mathtt b$, and~$\mathtt c$. 
Let us use $\ell\p m$ to denote the label of message $m$.
Assuming $\mathtt p$ trusts $\mathtt a,\mathtt b,$ and $\mathtt c$ to send these messages, the least restrictive 
guarantee that can be made regarding their availability is that 
$A\p{\ell\p{m_{\mathtt a}}} = \cb{\mathtt p \avail \mathtt a}, A\p{\ell\p{m_{\mathtt b}}} = \cb{\mathtt p \avail \mathtt b},$ and 
$ A\p{\ell\p{m_{\mathtt c}}} = \cb{\mathtt p \avail \mathtt c}$. 
Suppose the message synthesizer $\pi_\mathit{fastest}$ listens for these messages
and returns whichever message arrives first. 
This synthesizer produces a value that is more available (has less restrictive availability) 
than any one of the messages, because the value is unavailable only if
all three of $\mathtt a$, $\mathtt b$, and $\mathtt c$ are unavailable:

\ifreport
	\[
	A\p{\ell\p{\pi_\mathit{fastest}}}=
	A\p{\ell\p{m_{\mathtt a}}}\meet A\p{\ell\p{m_{\mathtt b}}}\meet
	A\p{\ell\p{m_{\mathtt c}}}
	=  \cb{\mathtt p \avail \p{\mathtt a \land \mathtt b \land\mathtt c}}
	\]
\else
	\begin{multline*}
	A\p{\ell\p{\pi_\mathit{fastest}}}=
	A\p{\ell\p{m_{\mathtt a}}}\meet A\p{\ell\p{m_{\mathtt b}}}\meet
	A\p{\ell\p{m_{\mathtt c}}}
	= \\ \cb{\mathtt p \avail \p{\mathtt a \land \mathtt b \land\mathtt c}}
	\end{multline*}
\fi

However, this synthesizer allows $\mathtt a$, $\mathtt b$, or $\mathtt c$ to affect its value, so its 
integrity is correspondingly lowered:

\ifreport
	\[
	I\p{\ell\p{\pi_\mathit{fastest}}} = 
	I\p{\ell\p{m_{\mathtt a}}}\join  I\p{\ell\p{m_{\mathtt b}}}  \join
	I\p{\ell\p{m_{\mathtt c}}}
	=  \cb{\mathtt p \integ \p{\mathtt a \lor \mathtt b \lor\mathtt c}}
	\]
\else
	\begin{multline*}
	I\p{\ell\p{\pi_\mathit{fastest}}} = 
	I\p{\ell\p{m_{\mathtt a}}}\join  I\p{\ell\p{m_{\mathtt b}}}  \join
	I\p{\ell\p{m_{\mathtt c}}}
	= \\ \cb{\mathtt p \integ \p{\mathtt a \lor \mathtt b \lor\mathtt c}}
	\end{multline*}
\fi

In contrast, we might have a different message synthesizer $\pi_\mathit{all}$ that 
returns a value only once it receives all three messages, and only if all three 
carry identical values. 
In this case, the results are available only if all the message senders are 
available (any one sender can render it unavailable), but the integrity of the 
result is much less restrictive: the result can be corrupted only
if all three messages were corrupted:

\ifreport
	\[
	A\p{\ell\p{\pi_\mathit{all}}} =
	A\p{\ell\p{m_{\mathtt a}}}\join A\p{\ell\p{m_{\mathtt b}}}  \join
	A\p{\ell\p{m_{\mathtt c}}} = 
	\cb{\mathtt p \avail (\mathtt a\lor \mathtt b\lor \mathtt c)}
	\]

	\[
	I\p{\ell\p{\pi_\mathit{all}}} =
	I\p{\ell\p{m_{\mathtt a}}}\meet  I\p{\ell\p{m_{\mathtt b}}}  \meet
	I\p{\ell\p{m_{\mathtt c}}} = 
	\cb{\mathtt p \integ (\mathtt a\land \mathtt b\land \mathtt c)}
	\]
\else
	\begin{multline*}
	A\p{\ell\p{\pi_\mathit{all}}} =
	A\p{\ell\p{m_{\mathtt a}}}\join A\p{\ell\p{m_{\mathtt b}}}  \join
	A\p{\ell\p{m_{\mathtt c}}} = \\
	\cb{\mathtt p \avail (\mathtt a\lor \mathtt b\lor \mathtt c)}
	\end{multline*}
	\vspace{-2em}
	\begin{multline*}
	I\p{\ell\p{\pi_\mathit{all}}} =
	I\p{\ell\p{m_{\mathtt a}}}\meet  I\p{\ell\p{m_{\mathtt b}}}  \meet
	I\p{\ell\p{m_{\mathtt c}}} = \\
	\cb{\mathtt p \integ (\mathtt a\land \mathtt b\land \mathtt c)}
	\end{multline*}
\fi

% \[\small
% \begin{array}{rcccl}
% A\p{\ell\p{\pi_\mathit{all}}} &=&
% A\p{\ell\p{m_{\mathtt a}}}\join A\p{\ell\p{m_{\mathtt b}}}  \join
% A\p{\ell\p{m_{\mathtt c}}} &=&
% \mathtt p \avail (\mathtt a\lor \mathtt b\lor \mathtt c)
% \\
% I\p{\ell\p{\pi_\mathit{all}}} &=&
% I\p{\ell\p{m_{\mathtt a}}}\meet  I\p{\ell\p{m_{\mathtt b}}}  \meet
% I\p{\ell\p{m_{\mathtt c}}} &=&
% \mathtt p \integ (\mathtt a\land \mathtt b\land \mathtt c)
% \end{array}
% \]

The availability constraint poses an interesting problem here: sometimes we want
to know if an availability constraint is met. 
For example, we might want to know if $\pi_\mathit{all}$ received contradictory 
messages. We follow Zheng~\cite{zlt-thesis} by using the special value \tb{none}
to denote a detectably unavailable value.
%\ACM{Is this an unavailable value, or an available but corrupted
%value?  It seems like the latter, though I don't recall what Lantian
%said about it.}
%\ICS{I suppose here it's ``unavailable'' in the sense that it didn't meet the
%given availability requirements. I'm not sure exactly how to phrase it, either.}

%\cite{zlt-thesis} also introduces the message synthesizer for ``label 
%thresholds'' $LT\sqb l$, which returned a value if the set of messages 
%containing identical copies of that value together had an integrity label less 
%restrictive than integrity policy $l$. 
%That is to say that if $m_a$ and $m_b$ have value $v$, and $m_c$ had some other
%value $v^\prime$, $LT\sqb l$ would return $v$ if 
%$I\p{\ell\p{m_a}}\meet I\p{\ell\p{m_b}} \less_I l$, it would return $v^\prime$ 
%if $I\p{\ell\p{m_c}}\less_I l$, and \tb{none} otherwise. 
%Note that this does provide for the possibility in which $LT\sqb{l}$ can return
%multiple things. 
%In this case, it ought to return whichever value is possible first.

A consensus protocol is a form of message synthesizer. 
%\ACM{Not strictly, right? Each round is a message synthesizer but the
%thing as a whole is interactive, which message synthesizers cannot
%be. I think we should be more precise here.}
%\ICS{There's no rule in the definition provided that says a message synthesizer
%can't send and receive messages. In this case, you build a message synthesizer
%with a starting value, it computes, and then synthesizes a message output.
%Computing just happens to involve sending and receiving messages.}
It takes in messages, and synthesizes a consensus message,
accompanied by availability and integrity guarantees that
no one message could have. 

% In a consensus protocol, we assume each principal $p$ begins with some value 
% $v_p$ such that that principal trusts itself to write that value, and to make 
% it available.\RVR{Why talk about a consensus protocol here?}

\subsection{Comparison to survivor sets and failure-prone sets}
\label{survivorsets}
A different way to express the powers of an attacker is to use _failure-prone sets_
of principals.
\cite{survivor-sets,Junqueira2005,Malkhi97a}.
A non-blocking protocol is one in which
no failure of any subset of a failure-prone set prevents termination or 
progress.
The complements of failure-prone sets are _survivor sets_. 
Every failure of any subset of a failure prone set leaves at least one survivor 
set failure-free. 

We can therefore express a principal $p_{sys}$ that has the least restrictive 
integrity or availability the system can have, as the disjunction of each of 
the survivor sets, each represented by the conjunction of its 
members.

Similarly, we can express a principal $p_{attack}$ that has the most restrictive 
availability or integrity the attacker can't have, as the weakest compound 
principal such that it is impossible to partition the set of principals into
one partition whose conjunction implies $p_{sys}$, and another whose conjunction
implies $p_{attack}$. 
A survivor set survives iff the attacker cannot act for $p_{attack}$,
and so in the case of crash failures, a piece of information that must be
supplied by a survivor set to some principal $o$ has the availability 
$o\avail p_{attack}$, and in the case of Byzantine failures, the integrity 
$o\integ p_{attack}$.

For example, in a system of four participants $P=\cb{\mathtt a,\mathtt
b,\mathtt c,\mathtt d}$, any one of which might fail we have:

\ifreport
	\[
	\begin{array}{r c l}
	p_{sys}&=&\p{\mathtt a\land \mathtt b\land \mathtt c}\lor\p{\mathtt a\land \mathtt b\land \mathtt d}\lor
	      \p{\mathtt a\land \mathtt c\land \mathtt d}\lor\p{\mathtt b\land \mathtt c\land \mathtt d} \\
	p_{attack}&=&\p{\mathtt a\land \mathtt b}\lor\p{\mathtt a\land \mathtt c}\lor\p{\mathtt a\land \mathtt d}\lor
	     \p{\mathtt b\land \mathtt c}\lor \p{\mathtt b\land \mathtt d}\lor\p{\mathtt c\land \mathtt d}
	\end{array}
	\]
\else
	\[
	\begin{array}{r c l}
	p_{sys}&=&\p{\mathtt a\land \mathtt b\land \mathtt c}\lor\p{\mathtt a\land \mathtt b\land \mathtt d}\lor
	          \\&&
	      \p{\mathtt a\land \mathtt c\land \mathtt d}\lor\p{\mathtt b\land \mathtt c\land \mathtt d} \\
	p_{attack}&=&\p{\mathtt a\land \mathtt b}\lor\p{\mathtt a\land \mathtt c}\lor\p{\mathtt a\land \mathtt d}\lor
	          \\&&
	     \p{\mathtt b\land \mathtt c}\lor \p{\mathtt b\land \mathtt d}\lor\p{\mathtt c\land \mathtt d}
	\end{array}
	\]
\fi

Using this construction,
any protocol phrased in terms 
of survivor and failure-prone sets can be expressed in terms of labels instead. 
Since the condition that no more than $f$ failures occur can be converted to
survivor and failure-prone sets, it can also be expressed in terms of
labels.

There is also a certain kind of equivalence between labels and
survivor sets: any label can be expressed in terms of (for
each principal) two collections of survivor and failure-prone sets,
describing the crash and Byzantine failures that principal tolerates.

%Phrasing them in terms of collections of policies, and forming a lattice, allows some higher-order reasoning, as well as the power of the body of programming languages research featuring labels~\cite{ml-tosem,cm06,zm05,labw91,chong-thesis,sm-jsac,zznm02,zcmz03,zm14,zlt-thesis,jif}. % DISC reviewer 2, you wanted to know why use labels and not just survivor sets.

\section{Heterogeneous fast consensus}
\newcommand\definedas{~~\equiv~~}
\newcommand\bigmeet{\bigsqcap}

\newcommand{\underlyingconsensus}{\textit{underlying-consensus}\xspace}
\newcommand{\fastconsensus}{\textit{fast-consensus}\xspace}
\newcommand{\selectionfunction}{\textit{selection-function}\xspace}

\label{heterogeneous-consensus}
As an example of a protocol adapted for a heterogeneous trust environment, we 
present Heterogeneous Fast Consensus, a generalization of the Bosco Fast 
Consensus protocol~\cite{Song2008}.
Fast Consensus is a one-round protocol that can, in the best case,
decide on a consensus value in one communication step.
If it fails to decide, some underlying consensus is used;
this can be another round of Fast Consensus.
The desirable properties of a traditional consensus protocol~\cite{Bracha85,Song2008}
can be generalized for heterogeneous trust:

\noinbf{Agreement:}
\label{agreement}
If two gurus decide, then they decide the same value. 
Also, if a guru decides more than once, it decides the same value each time.

\noinbf{Unanimity:}
\label{unanimity}
If all correct participants have the same initial value $v$, a guru that 
decides must decide $v$.

\noinbf{Validity:}
\label{validity}
If a participant decides a value, and all participants are correct,
then that value was proposed by some participant. 

\noinbf{Progress:}
Under the assumption that underlying consensus
terminates, all gurus eventually decide.

\noinbf{Termination:}
% If {\underlyingconsensus}() recursively invokes another instance of
% {\fastconsensus}, then
Under the assumption that the network delivers concurrent messages in
random order, and underlying consensus is simply a recursive
invocation of another instance of Fast Consensus, all gurus eventually
decide with probability~1.
\smallskip

Algorithm~\ref{algorithm:pseudocode-fast-consensus} contains the
pseudo-code for Heterogeneous Fast Consensus, and
Appendix \ref{hetconsproofs} contains the proof of correctness.
Each participant broadcasts its starting value to each participant
(including itself). 
Once it can no longer be sure it will receive any more messages (given its 
failure assumptions), the participant looks over the values it has received.
If a set of identically-valued messages meets the high, \textit{decision}, 
threshold, that participant can decide that value.
What a participant does to \textit{decide} varies depending upon 
{\underlyingconsensus}(), but if {\underlyingconsensus}() is {\fastconsensus}(), it simply
broadcasts the decided value one last time:
\begin{tabbing}
XXX\=XXX\=\kill
\>$\mathit{decide}\p v : \{$ \\
\>\> $\tb{send message with value }v\tb{ to } \textrm{each participant};$ \\
\>\> $\tb{return} ~ v$ \\
\>\}
\end{tabbing}

If some value has a set of messages which meet a lower \textit{change} threshold, 
then the participant enters {\underlyingconsensus}() with that as its starting
value.
Otherwise, the participant picks a value $v$ from those received using some
{\selectionfunction}(), and invokes ${\underlyingconsensus}(v)$.

\begin{algorithm}
\SetAlgoLined
\SetKwProg{Fn}{Function}{:}{end}
\SetKw{none}{none}
\SetKw{send}{send message with value}
\SetKw{to}{to}
\SetKwProg{Ur}{Upon receipt of a message}{:}{end}
\Fn{fast-consensus($v_p$)}{
  
  \send $v_p$ \to each participant\;
  \Ur{}{
    $R \longleftarrow $the set of messages received thus far\;
    \If{sufficiently-available($R$)}{
      \ForAll{unique values $v$ in $m\in R$}{
        $S \leftarrow \cb{m |\p{\textrm{value of }m} = v \land m\in R}$\;
        \uIf{sufficient-to-decide($S$)}{
          \Return decide$\p v$\;
        }
        \ElseIf{sufficient-to-change$\p S$}{
          \Return {\underlyingconsensus}$\p v$
        }
      }
      \Return {\underlyingconsensus}(
	  					\rule{1in}{0in}{\selectionfunction}$\p R$)
    }
  }
}
\caption{
Pseudo-code for Fast Consensus \cite{Song2008}. 
The input is the starting value for a given participant. 
Functions \textit{sufficiently-available}(), \textit{sufficient-to-decide}, and 
{\selectionfunction} are discussed elsewhere. 
Function {\underlyingconsensus} is the consensus protocol to be used in the case that 
this one-round consensus fails to decide. 
It is assumed to take as input a participant's
starting value, and it can be another round of {\fastconsensus}(). 
This pseudo-code assumes a language mechanism for generating a list of expected 
messages for this particular consensus round.
}
\label{algorithm:pseudocode-fast-consensus}
\end{algorithm}

The function {\selectionfunction}()
varies depending upon the desired properties 
of the protocol and on {\underlyingconsensus}().
If {\underlyingconsensus}() is {\fastconsensus}(), the protocol may converge 
fastest if {\selectionfunction} always selects the first in some arbitrary but 
consistent ordering of the input values. However,
this deterministic strategy may permit a Byzantine attacker to 
prevent agreement in each round. 
It is therefore prudent to incorporate some randomness in {\selectionfunction}(), 
as in RS-Bosco \cite{Song2008}.

\subsection{Modeling simple homogeneous trust}

In the traditional case, in which all $n$ participants believe that any $c$ 
participants may crash, and any $b$ participants may fail in Byzantine fashion
(which traditionally includes crashing, so $c \geq b$), the thresholds are 
fairly straightforward \cite{Song2008}:
\begin{itemize}
\itemsep 1.5pt
\item $\textit{sufficiently-available}\p R \definedas \abs R \geq n-c$
\item $\textit{sufficient-to-decide}\p S \definedas \abs S > \frac{n+c}{2}+b$
\item $\textit{sufficient-to-change}\p S \definedas \abs S > \frac{n-c}{2}$
\end{itemize}
\noinbf{Requirements:} This protocol has similarly straightforward 
requirements on the values of $n$, $c$ and $b$. 
A participant can only be sure of 
receiving $n-c$ votes, and needs more than $\frac{n+c}{2}+b$ to decide anything,
so it is required that $n>3c + 2b$. 
If only crash failures are expected $\p{b=0}$, then 
this means $n>3c$, and if only Byzantine failures are expected $\p{b=c}$, then 
this means $n>5b$.
These bounds are known to be tight for single-communication-step consensus 
\cite{Song2008}.

\subsection{Heterogeneous Trust}
In an information flow setting, labels are assigned to each piece of 
information to describe properties such as availability and integrity, 
characterizing who might make that information unavailable
or affect its contents (section~\ref{ifc}). 
In this case, each message $m$ has a corresponding label $\ell\p m$.
A set of messages taken together can be used to synthesize a value with more
integrity or availability than any message alone (section~\ref{synthesizers}). 

Each participant may have a different idea of the possible failures.
A participant $p$ can phrase its requirement for when it believes it cannot
be certain of receiving any more messages as, for some availability
policy $A^p_{\textrm{sys}}$ unique to $p$,
the condition $\bigmeet_{m\in R}\,A\p{\ell\p m}\less A^p_{\textrm{sys}}$. 
Label $A^p_{\textrm{sys}}$ may be constructed as in section~\ref{survivorsets}.
As $A^p_{\textrm{sys}}$ should contain only policies owned by $p$, and the only portions
of the label on $m$ which $p$ must enforce are those owned by $p$, we can 
define:

$A_{\textrm{sys}} = \bigmeet_{p\in P}A^p_{\textrm{sys}}$, and can then write:

\[ \textit{sufficiently-available}\p R \definedas
    \bigmeet_{m\in R}\,A\p{\ell\p m}\less A_{\textrm{sys}}
\]

Similarly, a threshold $C^p$, composed of both availability and integrity 
policies, represents when a participant feels a set of messages should be
sufficient to force it to change its starting value to a particular value in
{\underlyingconsensus}. 
We define $C=\bigmeet_{p\in P}\,C^p$, and:
\[
\textit{sufficient-to-change}\p S \definedas \bigmeet_{m\in S}\,\ell\p m \less C
\]

Similarly, a threshold $D^p$, composed of both availability and integrity 
policies, represents when a participant feels a set of messages should be
sufficient to decide on a value carried by all of those messages.
We define $D=\bigmeet_{p\in P}D^p$, and:
\[
\textit{sufficient-to-decide}\p S \definedas
\bigmeet_{m\in S}\,\ell\p m \less D
\]

\subsection{Requirements}
\label{sec:requirements}

As in the case of simple homogeneous trust, heterogeneous trust
imposes requirements on the values of $A_{\textrm{attack}}$,
$A_{\textrm{sys}}$, $I_{\textrm{attack}}$, $I_{\textrm{sys}}$, $C$,
and $D$, as well as on the labels of
messages sent as part of the protocol.
%\RVR{I'm not sure if we know at this point what $A_a$, $I_a$, and $I_s$ are.  Can we remove these here or define them first?}
%\ICS{They're defined in the preceding paragraphs. That is the purpose of those paragraphs.}
%\RVR{I looked again at those paragraphs, but I don't see the definitions.}

Let $m^p_q$ designate a message from participant $p$ to participant~$q$.
Let $A^p_{\textrm{attack}}$ designate the most restrictive availability label 
participant $p$ believes an attacker can't block.
Let $I^p_{\textrm{attack}}$ designate the most restrictive integrity label 
participant $p$ believes an attacker can't influence.

From the perspective of a principal $p$, therefore, we can define
failure-prone sets for Byzantine and crash failures, called
Liar sets and Crash sets:
\[
L\textrm{ is a liar set }\Leftrightarrow
\p{\bigsqcap_{p\in P, q\in L}\cb{p\integ q}} ~ \notless ~ I^p_{\textrm{attack}}
\]
\[
H\textrm{ is a crash set }\Leftrightarrow
\p{\bigsqcap_{p\in P, q\in H}\cb{p\avail q}} ~ \notless ~ A^p_{\textrm{attack}}
\]
% \[
% \begin{array}{l	|	r}
% L\textrm{ is a liar set }\Leftrightarrow
% \p{\bigsqcap_{p\in P, q\in L}\cb{p\integ q}}\notless I^p_{\textrm{attack}}
% \ \	&\ \	
% H\textrm{ is a crash set }\Leftrightarrow
% \p{\bigsqcap_{p\in P, q\in H}\cb{p\avail q}}\notless A^p_{\textrm{attack}}
% \end{array}
% \]
It is possible, for example, to construct $A^p_{\textrm{sys}}$ given $A^p_{\textrm{attack}}$ from survivor 
sets: the crash sets' compliments.
This notation simplifies future expressions.

A participant may not be able to send a message if it is itself unavailable. 
Therefore, from the perspective of participant $q$, the availability of a 
message $m^p_q$ from participant $p$ is limited by this constraint:
\begin{equation}\label{eq:message-avail-limit}
\cb{q \avail p} \less A\p{\ell\p{ m^p_q}}
\end{equation}

Each participant collects received messages $R$ until the meet of
their labels of all received messages is no more restrictive than $A_{\textrm{sys}}$. 
This places a limit on \textit{viable} systems. 
For no participant should it ever be the case that the attacker (as perceived by
that participant) can prevent such a set of messages from arriving: 
\ifreport
	\begin{equation}\label{eq:message-set-limit}
	\forall q\in P.~
	\forall F\subseteq P.~
	\p{q\not\in F}\land \p{\bigsqcap_{f\in F}\cb{q\avail f}\notless A_{\textrm{attack}}}
	\Rightarrow 
	\bigsqcap_{p\in\p{P-F}}
	\p{A\p{\ell\p{m^p_q}}}
	\less A_{\textrm{sys}}
	\end{equation}
\else
	\begin{multline}\label{eq:message-set-limit}
	\forall q\in P.~
	\forall F\subseteq P.~
	\p{q\not\in F}\land \p{\bigsqcap_{f\in F}\cb{q\avail f}\notless A_{\textrm{attack}}}
	\Rightarrow \\
	\bigsqcap_{p\in\p{P-F}}
	\p{A\p{\ell\p{m^p_q}}}
	\less A_{\textrm{sys}}
	\end{multline}
\fi
If {\underlyingconsensus}() is {\fastconsensus}(), no participant should progress
to the next round unless it can complete that round with certainty.
This condition implies that if messages from a set of participants messages can propel $p$ to the
next round, but not $q$, then $p$ must be able to progress with that set,
without $q$. 
\ifreport
	\begin{equation}\label{eq:progress}
	\begin{array}{c}
		\forall r,q\in P.~
		\forall R\subseteq P.
		\\
		\p{\p{\bigsqcap_{p\in S}
		A\p{\ell\p{m^p_r}}
		\less A_{\textrm{sys}}}
		\land
		\p{
		\bigsqcap_{p\in S}
		A\p{\ell\p{m^p_q}}
		\notless A_{\textrm{sys}}}}
		\\
		\Downarrow
		\\
		\p{\bigsqcap_{p\in S-\cb q}
		A\p{\ell\p{m^p_r}}
		\less A_{\textrm{sys}}
		}
	\end{array}
	\end{equation}
\else
	\begin{multline}\label{eq:progress}
	\forall r,q\in P.~
	\forall R\subseteq P.~
	\bigsqcap_{p\in S}
	A\p{\ell\p{m^p_r}}
	\less A_{\textrm{sys}}
	\Rightarrow \\
	\p{
	\bigsqcap_{p\in S}
	A\p{\ell\p{m^p_q}}
	\notless A_{\textrm{sys}}
	\Rightarrow
	\bigsqcap_{p\in S-\cb q}
	A\p{\ell\p{m^p_r}}
	\less A_{\textrm{sys}}
	}
	\end{multline}
\fi

Another requirement is that no set of failures a participant believes might
happen should prevent that participant from deciding. 
Additionally, any set of messages that makes a correct participant decide
should also be sufficient to dictate that participant's value in 
{\underlyingconsensus}.
\begin{align}\label{eq:threshold-order}
\p{A_{\textrm{sys}}\meet \cb{\top \integ \top}} \less D \less C
\end{align}

If a participant $p$ decides, it is useful to talk about \textit{decider} sets and their
complements, \textit{wrong sets}.
A decider set is any set of principals whose messages can make $p$ decide.
In much the same way that we might construct $A_{\textrm{attack}}$ and \textit{crash sets} from $A_{\textrm{sys}}$,
we can construct $W$ and \textit{wrong sets} from $D$.
\[
E\textrm{ is a decider set }\Leftrightarrow
\bigsqcap_{e\in E}\ell\p{m^e_p}\less D
\]
\[
G\textrm{ is a wrong set }\Leftrightarrow
\bigsqcap_{g\in G}\ell\p{m^g_p}\notless W
\]
% \[
% \begin{array}{l	|	r}
% E\textrm{ is a decider set }\Leftrightarrow
% \bigsqcap_{e\in E}\ell\p{m^e_p}\less D
% \ \	&\ \	
% G\textrm{ is a wrong set }\Leftrightarrow
% \bigsqcap_{g\in G}\ell\p{m^g_p}\notless W
% \end{array}
% \]
A wrong set is any set of participants who broadcast, or may have broadcast, a 
value other than the one $p$ decided during the round in which $p$ decided.
The label $W$ dictates the least restrictive availability and
integrity that no wrong set may have.

\begin{figure*}
\centering
\[
\begin{array}{rlcl}
& A^{\mathtt a}_{\textrm{sys}}	&	=	&\cb{{\mathtt a}		\avail	
\p{{\mathtt a} \land {\mathtt c} \land {\mathtt d} \land {\mathtt e}}}\\

\forall p\in \cb{\mathtt b,\mathtt c,\mathtt d,\mathtt e}.&
A^{p}_{\textrm{sys}}&	=	&\cb{p		\integ		p\land
\p{\bigvee_{x,y,z\in \cb{\mathtt b,\mathtt c,\mathtt d,\mathtt e}}\p{x\land y \land z}}}\\

\forall p\in P.&
C^{p}&	=	&\cb{p		\integ		
\bigvee_{x,y\in \cb{\mathtt b,\mathtt c,\mathtt d,\mathtt e}}\p{x\land y}}
\\

\forall p\in P.&
D^{p}&	=	&\cb{p		\integ		
\bigvee_{x,y,z\in \cb{\mathtt b,\mathtt c,\mathtt d,\mathtt e}}\p{x\land y \land z}}
\\

\forall p,q\in P.&
A\p{\ell\p{m^{\mathtt q}_p}}	&	=	&	\cb{p		\avail	
q}
\\

\forall p,q\in P.&
I\p{\ell\p{m^{\mathtt q}_p}}	&	=	&	\cb{p		\integ	q}
\end{array}
\]
\caption{Using {\fastconsensus} as {\underlyingconsensus}, and a {\selectionfunction} of their 
choice (e.g., random selection), the 
participants in the example can execute fast consensus with these threshold values and labels.}
\label{fig:examplelabels}
\end{figure*}

Assuming {\underlyingconsensus} has unanimity, Fast 
Consensus assures agreement by ensuring that if one guru
% (see \ref{gurus})
decides, then all correct participants who have not yet
decided enter {\underlyingconsensus} with the value decided.
Fast Consensus therefore has some requirements pertaining to $C$
and $D$.

If $p$ is a guru, then none of $p$'s perceived wrong sets, combined with
any of $p$'s perceived liar sets, should be able to change anyone's vote.
Otherwise, a set of liars combined with a set of participants $p$ is aware
may have broadcast something other than it decided could prevent correct
participants from entering {\underlyingconsensus} with unanimous values.
\ifreport
	\begin{equation}
	\label{eq:wrong-liar-change}
	\forall p,q.~
	\p{\p{\bigsqcap_{l\in L}\cb{p\integ l}}\notless I^p_{\textrm{attack}}
	\land
	\p{\bigsqcap_{h\in H}\cb{p\avail h}}\notless W}
	\Rightarrow 
	\p{\bigsqcap_{x \in L\cup H}\ell\p{m^x_q}}\notless C
	\end{equation}
\else
	\begin{multline}
	\label{eq:wrong-liar-change}
	\forall p,q.\p{\bigsqcap_{l\in L}\cb{p\integ l}}\notless I^p_{\textrm{attack}}
	\Rightarrow
	\p{\bigsqcap_{h\in H}\cb{p\avail h}}\notless W
	\Rightarrow \\
	\p{\bigsqcap_{x \in L\cup H}\ell\p{m^x_q}}\notless C
	\end{multline}
\fi

We also require that if one guru decides, it must be impossible for any other 
correct participant not to enter {\underlyingconsensus} with the decided value.
This means that for any group of liars $L$, and any group of crashers $H$, and
any additional group $J$, if messages from $L\cup H\cup J$ make a guru decide, 
then messages from $J$, being the only ones guaranteed to get through to the 
other participants, must make those participants enter {\underlyingconsensus}
with the decided value.
\ifreport
	\begin{equation}\label{eq:decide-change}
	\begin{array}{c}
	\forall p,q\in P.\forall L,H,J\subseteq P. \\
	\p{
	\p{\p{\bigsqcap_{l\in L}\cb{p \integ l}}\notless I^p_{\textrm{attack}} }\land
	\p{\p{\bigsqcap_{h\in H}\cb{q \avail     h}}\notless A^q_{\textrm{attack}} }\land
	\p{\p{\bigsqcap_{x\in L\cup H\cup J}\ell\p{m^x_p}}\less D}}
	\\ \ \\
	 \Downarrow
	 \\ \ \\
	\p{\bigsqcap_{j\in J}\ell\p{m^j_q}}\less C
	\end{array}
	\end{equation}
\else
	\begin{multline}\label{eq:decide-change}
	\forall p,q\in P.\forall L,H,J\subseteq P. \\
	\p{\bigsqcap_{l\in L}\cb{p \integ l}}\notless I^p_{\textrm{attack}} \Rightarrow
	\p{\bigsqcap_{h\in H}\cb{q \avail     h}}\notless A^q_{\textrm{attack}} \Rightarrow \\
	\p{\bigsqcap_{x\in L\cup H\cup J}\ell\p{m^x_p}}\less D   \Rightarrow
	\p{\bigsqcap_{j\in J}\ell\p{m^j_q}}\less C
	\end{multline}
\fi

%Right now, Heterogeneous Consensus is our flagship example. 

%State and explain requirements

% TERMINATION
%\ICS{I'm not sure if I can prove this termination with probability 1 exactly 
%right. I need to specify ``fair links.''}
%\subsection{Termination}
%\subsubsection{Definition}
%All gurus eventually decide.
%
%
%\subsubsection{Proof}
%If {\underlyingconsensus} guarantees termination, then progress guarantees that
%{\fastconsensus} guarantees termination.
%
%If unerlying-consensus is fast consensus, then termination is guaranteed with
%probability 1 if the network has ``fair links,'' and {\selectionfunction} has
%some probability of returning each entry from its input set.
%This is because there will eventually occur two rounds in a row in which all
%correct participants observe messages from other correct participants first, and
%if any of them meet a change threshold for some value, those who execute 
%{\selectionfunction} will choose that value as well (or everyone chooses
%the same value via {\selectionfunction}). 
%Therefore, in the second round, all gurus will decide on that value. 

%proofs should probably get shifted to appendix, but we should mention relevant properties here.

%example
\subsection{Example}
Returning to the example introduced in section \ref{example-intro}, we can now 
synthesize a consensus protocol for Alice, Bob, Carol, Dave, and Eve. 
For brevity, the letters $\mathtt a$--$\mathtt e$ are used to represent the five
participants.

\noinbf{Alice}
believes that Bob may fail in a Byzantine fashion (lose integrity).
She does not believe any other failures may occur.

\noinbf{Bob, Carol, Dave, and Eve}
each believe Alice can fail in a Byzantine fashion (lose integrity), and
believe that at most one other participant may crash (lose
availability) as well.

These trust assumptions are captured by the following labels:
\[
\begin{array}{rl}
A^{\mathtt a}_{\textrm{attack}}	&	=		\cb{\mathtt a		\avail		\mathtt a \lor \mathtt c \lor \mathtt d \lor \mathtt e }\\
I^{\mathtt a}_{\textrm{attack}}	&	=		\cb{\mathtt a		\integ		\mathtt a \lor \mathtt c \lor \mathtt d \lor \mathtt e}\\

\forall p \in \cb{\mathtt b, \mathtt c, \mathtt d, \mathtt e}.
A^p_{\textrm{attack}}	&	=		\cb{p		\avail		p \lor \p{\bigvee_{q,r \in \cb{\mathtt b, \mathtt c, \mathtt d, \mathtt e}}\p{q\land r}}}
\\
\forall p \in \cb{\mathtt b, \mathtt c, \mathtt d, \mathtt e}.
I^{p}_{\textrm{attack}}	&	=		\cb{p		\integ		\mathtt b \lor \mathtt c \lor \mathtt d \lor \mathtt e}\\
\end{array}
\]

\noinbf{Solution:}
Search of the space of threshold labels ($A_{\textrm{sys}}, C, D$) reveals that there are 
indeed thresholds meeting the participants' requirements, as well as the 
requirements of Fast Consensus.
Using {\fastconsensus} as {\underlyingconsensus}, and a 
{\selectionfunction} of their 
choice (from a theoretical perspective, random selection works), they 
can execute Fast Consensus with the threshold values and 
labels found in Figure~\ref{fig:examplelabels}.

One counterintuitive insight provided by our analysis
%, and by deriving our protocol
%instance using the requirements derived heretofore,
is that there are occasions
in which Alice listens to Bob, despite the fact that she does not trust him at all.
From the label analysis, this falls out from $C^{\mathtt a}$ and $D^{\mathtt a}$, defined
in Figure~\ref{fig:examplelabels}, which can be shown to satisfy the threshold 
requirements.

Intuitively, the logic is this:
If Alice is correct, and Bob is Byzantine, and so everyone else is a chump, then it doesn't matter what Alice decides, so long as she does decide.
If, on the other hand, Bob is only crash-failure prone, then Alice can't decide having heard only from, for example, Carol and Dave, because Carol and Dave may have heard two votes from Eve and Bob, different from what they voted for, and change their votes next round.
As a result, Alice would have decided something different from what the others decide, despite no one having been wrong (or failing). 
Therefore, Alice must also wait to hear from Bob or Eve, to ensure that in the event that Bob is only crash-prone, Carol, Dave, and Eve will decide the same as what Alice decides.

\noinbf{Evaluation:}
\begin{figure}
\centering
Speed of Decision (No Failures)
\ifreport
	\includegraphics[width=\textwidth]{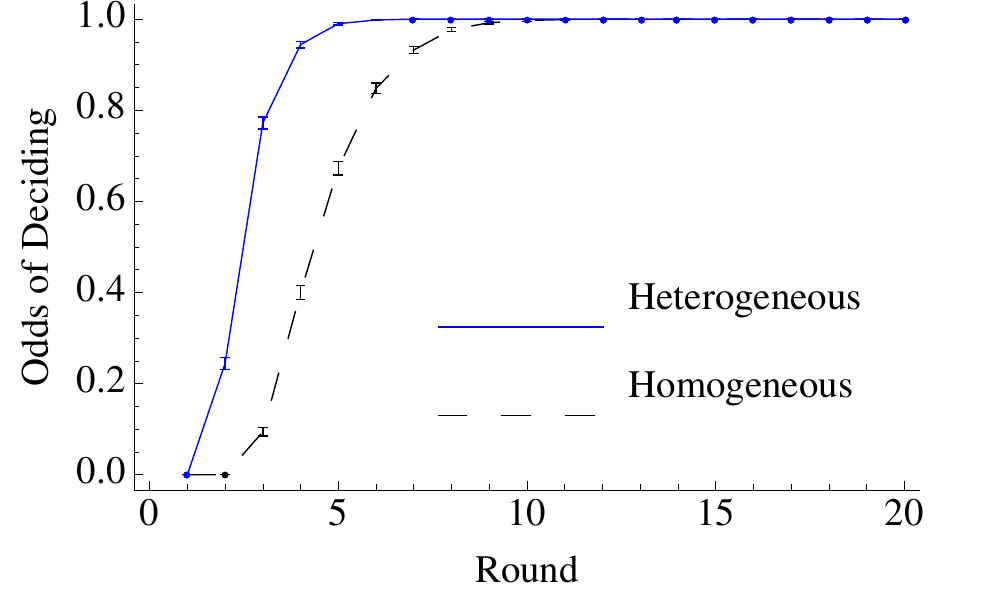}
\else
	\includegraphics[width=0.45\textwidth]{decisionspeedplot.pdf}
\fi
\caption{The probability of a participant deciding in each round, mean over
 1000 samples, in our 5-participant Heterogeneous Fast Consensus protocol, and
a traditional 9-participant Bosco protocol. 
Standard error bars shown.}
\label{fig:speeddecision}
\end{figure}
\begin{figure}
\centering
Speed of Decision (With Failures)
\ifreport
	\includegraphics[width=\textwidth]{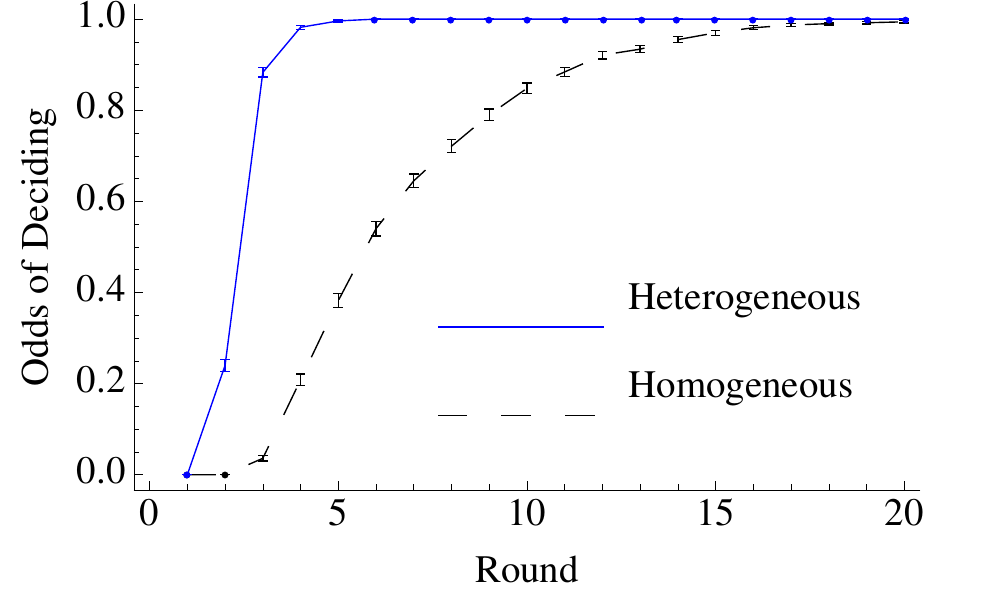}
\else
	\includegraphics[width=0.45\textwidth]{decisionspeedfailuresplot.pdf}
\fi
\caption{The probability of a participant deciding in each round, mean over
 1000 samples, in our 5-participant Heterogeneous Fast Consensus protocol, and
a traditional 9-participant Bosco protocol. In this case, Alice is Byzantine, and Bob has crashed.
Standard error bars shown.}
\label{fig:speeddecisionfailures}
\end{figure}
In this trust configuration, Eve tolerates the Byzantine failure of Alice and 
the simultaneous {\crash} failure of Bob. 
Therefore, traditional (homogeneous) Bosco would have to tolerate one Byzantine
failure and one \emph{additional} {\crash} failure, requiring a total of 9 
participants.
Already Heterogeneous Fast Consensus has a clear advantage for this
scenario: it requires only
5 participants to tolerate this trust configuration, whereas
traditional Bosco requires recruiting at least 4 more trustworthy participants, who
also slow the system down.
We simulated 1000 instances in which 9 participants participated in Bosco 
tolerating 1 Byzantine failure and 2 total {\crash} failures. 
The network delivered messages in each round in an order drawn uniformly at
random from all possible orderings. 
The {\selectionfunction}() used chose a value uniformly at random from the 
set of messages received. 

We also ran the same simulation for our Heterogeneous Fast Consensus
implementation, and for each calculated the mean (with standard error)
probability of a participant deciding after each round. 
All participants began with different values.

For the case in which no failures occurred, the results are in figure~\ref{fig:speeddecision}.
Not only did Heterogeneous Fast Consensus converge quickly, in a median of 3 
rounds, but it converged much faster than Homogeneous Bosco, in a
median time of 5 rounds. 
The gap was even wider in the 95th percentile, where Heterogeneous Fast Consensus 
took 5 rounds, and Homogeneous Bosco took 8.

For the case in which Alice has failed in a Byzantine fashion (specifically, she proposes a new, never-before-seen value each round), and Bob has crashed, the difference is even greater.
With the reduced contention of fewer active participants, Heterogeneous Fast Consensus converges even faster, the 95th percentile deciding by round 4, while the homogeneous case takes until round 6 for the median to decide, and  round 12 before the 95th percentile decided. The full results are in figure~\ref{fig:speeddecisionfailures}.

These results suggest that customizing the protocol to the
heterogeneous trust configuration yields clear advantages in both
resource requirements and speed.

\noinbf{Determining threshold labels}

One challenge of the Heterogeneous Fast Consensus protocol is finding
appropriate threshold labels to satisfy the requirements of
section~\ref{sec:requirements}. This is an offline computation, so
performance is not critical. We expressed the requirements using
quantifier-free bitvector logic (QF\_BV), and used the Z3 SMT solver
to find solutions. Generating four- to six-participant 
protocols took a few minutes for each protocol. The time to generate
protocols increases as the number of participants increases, but one
does not normally generate consensus protocols for very large numbers
of participants in any case.

We have made the search script available at
{\small
\url{https://www.dropbox.com/s/akv957fmrqsn803/pysmt.zip?dl=0}}.

Generalized heterogeneous fault-tolerant protocols will have,
in general, parameters specific to the needs of their participants.
For example, heterogeneous fast consensus has the thresholds $A_{\textrm{sys}}$, $C$, and 
$D$, which must be crafted to fit the requirements of the protocol 
and the specific distrust of the participants.
The exact nature of the complexity of such a search is unclear.  For
any given set of labels, it is easy (a polynomial-time computation) to
check that they meet these constraints.  The problem is therefore in
\textsc{np}, but the precise hardness of the problem remains future work.

\subsection{Very Fast Consensus}
This construction of fast consensus suggests another way to exploit
heterogeneous trust.
Suppose a client wishes to submit a request to a group of servers, from which
it requires an answer upon which the servers must reach consensus.
Suppose this operation is extremely latency-sensitive, and the goal is best-case performance
of one communication to the servers, no communication time between servers, and 
one communication to the client.
This construction of fast consensus extends to cover this extremely
latency-sensitive case simply by adding the client to the set of participants, as
a participant whom the servers don't trust at all.
The client thus receives messages as a participant, and can set its own 
thresholds for when it is satisfied consensus has been reached (subject, of 
course, to the protocol's requirements). 
This is not efficient in terms of bandwidth, but highly efficient in terms 
of latency, as the client need not wait for any communication between servers 
before servers send it a response. 

%\subsection{Other distributed protocols}

%\ACM{Do we want to mention that we have applied this approach to other
%protocols like OARcast, even if we don't have space to explain them? I
%think that would strengthen the argument even if it has the flavor of
%'just trust us'}
%We may choose to try and squeeze in Nyssiad or something later, but space is very limited.

\section{OARcast and Nysiad}
To further illustrate the utility of our approach, we present a second generalization of 
a distributed protocol, using the same techniques based on 
information flow analysis.
In this case, we generalize Nysiad, an algorithm for converting {\crash}-tolerant 
state-machine systems (\cite{Schneider90}) to Byzantine-tolerant ones~\cite{Ho2008,Ho2007}.
In particular, the conversion process allows a faulty ``sender'' participant to interfere with
the availability, but not the integrity, of a message.

\subsection{Ordered Asynchronous Reliable Broadcast}
At its core, Nysiad is built around Ordered Asynchronous Reliable 
broadcast (OARcast), a protocol that is useful in its own right.
OARcast, as presented in \cite{Ho2008,Ho2007}, works
as follows.

There exists a set of participants known as _echoers_.  One special
echoer, the _designated sender_, may wish to broadcast a message $m$
such that all other echoers receive it.  The goal of the protocol is
that all messages broadcast should arrive in the same order at all
nodes: the order in which they were sent.

Each message is assumed to be signed by both its author and its sender
(which may not be the author if the message is relayed through an echoer).
Each message is also assumed to contain a sequence number, assigned sequentially
by the author.

Each echoer will echo any new message (that it's not seen before) from
the designated sender to all other echoers.  Any echoer that receives
two messages signed by the designated sender, and containing the same
sequence number, that have different values, ceases operation.

An echoer _delivers_ a message (that is, produces a value) when it has delivered messages with
all lesser sequence numbers, and the set of identical messages its received for this
sequence number meets a condition.
This condition can be expressed with label threshold synthesizers, instead of 
waiting for a specific number of identical messages (See section ~\ref{synthesizers}).
The basic requirement, that no set of messages from echoers should allow two 
guru echoers to deliver different messages, remains the same.
Specifically, if each echoer "e" has an integrity value $I^{\mathtt e}_a$, 
defined to be strictly less restrictive than the integrity of any attacker it tolerates,
(expressed $\sqsubset$, meaning $x \sqsubset y \Leftrightarrow x \less y \land y \notless x$),
%\ACM{the use of $\sqsubset$ makes me nervous. Usually it has been
%wrong when we have considered using it before}
%\ICS{I do not know any other way of expressing this concept elegantly. If you do, change it.}
and has some integrity threshold $T_{\mathtt e}$ for delivering a message,
the following condition must hold to ensure gurus never deliver different messages:
\ACM{for what, exactly?}
\ICS{I don't know how to explain it much more clearly than ``The basic requirement, that no set of messages from echoers should allow two guru echoers to deliver different messages, remains the same.''}

\ifreport
	\begin{equation}
	\begin{array}{c}
	\forall \mathtt e,\mathtt {e^\prime}\in P. 
	\forall A,B,C\subseteq P: B \cap C = \cb{}. \\
	\p{I^{\mathtt e}_a \sqsubset \bigsqcap_{\mathtt q\in A}{I\p{\ell\p{m^{\mathtt q}_{\mathtt e}}}}}
	\land
	\p{{\bigsqcap_{\mathtt q\in \p{A\cup B}}{I\p{\ell\p{m^{\mathtt q}_{\mathtt e}}}}}\less T_{\mathtt e}}
	\\
	\Downarrow
	\\
	\p{{\bigsqcap_{\mathtt q\in \p{A\cup C}}{I\p{\ell\p{m^{\mathtt q}_{\mathtt {e^\prime}}}}}}\notless T_{\mathtt e^\prime}}
	\end{array}
	\end{equation}
\else
	\begin{multline}
	\forall \mathtt e,\mathtt {e^\prime}\in P. 
	\forall A,B,C\subseteq P: B \cap C = \cb{}. \\
	{I^{\mathtt e}_a \sqsubset \bigsqcap_{\mathtt q\in A}{I\p{\ell\p{m^{\mathtt q}_{\mathtt e}}}}}
	\land
	{{\bigsqcap_{\mathtt q\in \p{A\cup B}}{I\p{\ell\p{m^{\mathtt q}_{\mathtt e}}}}}\less T_{\mathtt e}}
	\\
	\Downarrow
	\\
	{{\bigsqcap_{\mathtt q\in \p{A\cup C}}{I\p{\ell\p{m^{\mathtt q}_{\mathtt {e^\prime}}}}}}\notless T_{\mathtt e^\prime}}
	\end{multline}
\fi

The proof of correctness proceeds exactly as in \cite{Ho2008,Ho2007}. Concerning safety properties, it should be clear
that:
\begin{itemize}
\item No two gurus can deliver different values for the same message sequence number.
\item If the designated sender is correct, then no correct participant (even a chump) can simulate delivery of messages in any order other than that provided by the designated sender.
\end{itemize}

\noindent For liveness, however, we need a guarantee
that:
\begin{itemize}
\item A guru will always deliver any message sent by a correctly functioning sender, so long as no attacker can compromise a set of echoers such that the remainder is insufficient for the guru. Formally:
%\ACM{I assume that the following formula captures this, but it's not
%obvious}
%\ICS{I hope it's obvious now.}
\end{itemize}
\[
\forall \mathtt e \in P.
\forall A \cup B = P.
{I^{\mathtt e}_a \sqsubset \bigsqcap_{\mathtt q\in A}{I\p{\ell\p{m^{\mathtt q}_{\mathtt e}}}}}
\Rightarrow
  {\bigsqcap_{\mathtt q\in \p{B}}{I\p{\ell\p{m^{\mathtt q}_{\mathtt {e}}}}}}\less T_e
\]
With this additional requirement, no attacker can prevent a guru from reaching its delivery conditions.

The integrity of the delivered message cannot therefore only be interfered with by the echoer's perceived attacker.\footnote{Integrity, as far as OARcast is concerned, is a property of uniform and guaranteed message delivery, 
and not content, which may have additional constraints.}
In other words, the integrity of the delivered message is $I^{\mathtt e}_a$.
Availability is limited by the sender, as well as any conditions under
which integrity is violated, calculated as follows:
\[
A_{\mathtt s}\meet
\bigsqcup_{M\subseteq \cb{m^{\mathtt p}_{\mathtt e}\left| p\in P  \land \meet_{m\in M}I\p{\ell\p m} \less I^{\mathtt e}_a\right.}}
\p{\bigsqcap_{m\in M} A\p{\ell\p{m}}}
\]
where $A_{\mathtt s}$ is the availability of the designated sender.

\subsection{Nysiad}

Nysiad, a translation mechanism from arbitrary crash-tolerant protocols to 
Byzantine tolerant ones, also presented in \cite{Ho2008,Ho2007}, can be 
performed using the heterogeneous OARCast.

The idea of Nysiad is to take any crash-tolerant system consisting of a collection of deterministic state machines, and simulate it on each of a group of participants.
For each state machine in the original system, the participants form an OARcast, and one participant is designated as the sender for that machine.
Each participant simulates the delivery of a message to each simulated state machine only if it has itself calculated an identical simulated message sent to that simulated machine, and it has received an identical message via that machine's OARcast. 
In this way, all gurus simulate identical executions, with identical message delivery ordering.
OARcast ensures that a failed participant cannot force two gurus to perceive message delivery in different orders, and the requirement that each participant derive the simulated messages themselves ensures that no failed participant can force another participant to deliver an incorrect value. 
The availability of simulated messages is thus limited by the availability  of the simulated sender machine's designated sender participant, as well as the availability of the sender machine's OARcast. 
The integrity of simulated messages (which really is all in the ordering) is limited only by the integrity of the sender machine's OARcast, and not the designated sender participant.
Additionally, all information simulated on a given participant is limited by that participant's availability and integrity. 

It is notable that, in the $f$-failure tolerant case, a Nysiad conversion of a $3f+1$ {\crash} tolerant Bosco instance (with a deterministic selection function) is a $3f+1$ Byzantine tolerant consensus protocol with best-case two message sends from proposal to decision, putting it on par (by this very simple metric), with Fast Byzantine Paxos~\cite{osdi99,fastpaxos}. 
A generalized version of Heterogeneous Fast Consensus can be likewise constructed for specific use cases.
\ACM{Not clear what the punchline is here. Do we have any actual
Nysiad results to report?}
\ICS{Well, we don't. I just like noting the consensus case. I hope the sentence below serves as a ``punchline.''}

As well as providing insightful cases in heterogeneous trust reasoning, the Nysiad algorithm, already a useful tool in constructing Byzantine-tolerant protocols, generalizes into a useful tool in heterogeneous trust based algorithms. 
%As the details of such an example would be by necessity complex and tedious, we do not attempt to present one here.

%\section{Results}
%\input{results}
%\section{Discussion}
%\input{discussion}
%\section{Related Work}
%\input{related}
\section{Future work}
The tools, techniques, and examples in this work are meant to provide
a framework for reasoning about and constructing fault-tolerant
distributed protocols.
We hope that protocol designers will expand on this approach to
to develop novel protocols.
Heterogeneous Fast Consensus is both a novel protocol and a
useful example of applying information-flow techniques to fault
tolerance. More efficient methods for synthesizing threshold
labels remain desirable.

The holy grail for heterogeneous trust would be
a procedure for transforming any
existing fault-tolerant protocol into a generalized version that
exploits heterogeneous trust. Such a procedure would require a
way to automatically derive necessary requirements on the trust
configuration and the protocol instances. While it is clear that such a
procedure will not always be computable (a desirable property might be
termination, and it is impossible to compute requirements for
termination in general~\cite{boyer1984}), it may be feasible for
useful cases.

The heterogeneous trust model of failure is extremely rich, but it
does not take into account notions of self-interest, and so there is
room for complementary work integrating game theory and selfish
participants into this richer space~\cite{bar-sosp,abraham2006}.  One
might envision, for example, protocols in which participants can
derive the specific implementations of an algorithm in which they take
part (as opposed to centrally determining this beforehand) using
knowledge of the trust configurations, and the belief that others will
derive their implementations selfishly.

Finally, we have deliberately ignored confidentiality in this work,
but confidentiality is also conducive to analysis using static
information-flow methods~\cite{sm-jsac}. Taking confidentiality into
account is likely to add additional constraints to protocol design.

\section{Conclusion}
% I've noticed that the DISC 2013 papers I've looked at don't seem to have
% conclusions. This seems odd to me.

In our increasingly complex, interconnected world, under varied and changing
threats and system models, it is critical to design systems that can operate in
environments where participants make differing trust assumptions about
the availability and integrity of information and of other participants.
We propose the use of information-flow labels describing integrity and
availability as a way to express those requirements and situations in a general
manner, and to provide a rigorous framework for reasoning about protocols
using heterogeneous trust. 
Our generalization of the Bosco Fast Consensus 
protocol \cite{Song2008}, developed with this methodology, is
capable of tolerating trust configurations for which
traditional fast consensus fails, or would be dramatically less efficient.
Properties such as Agreement, Unanimity, Validity, and Termination can be
generalized for the heterogeneous case, in which some but not all correct
participants make incorrect assumptions about failure. 

Likewise, our generalizations of OARcast and Nysiad~\cite{Ho2008,Ho2007}, 
and even basic message synthesizers may serve as useful tools and
building blocks in the development of future protocols that use heterogeneous trust.
The analysis of these example algorithms should serve to help others 
gain insight in future endeavors.

We expect that
our new approach will be useful for generalizing other fault tolerant
protocols to a heterogeneous trust environment, and we hope it will
lead to more efficient ways to build trustworthy systems.

%so, that's what we did . . . someone (maybe us) should build upon it, and make systems that deal with heterogenous trust . . .

\newpage
\appendix
%\section{Heterogeneous Consensus Examples}
%\input{hetconex}
\section{Principal and Lattice Formalisms}
\label{latticeappendix}
\subsection{Compound Principals and $\actsfor$}
The rules for reasoning about
 compound principals:

\ifreport
	\[
	\begin{array}{c}
	\begin{array}{c|c|c|c|c}
	(p_1 \land p_2) \actsfor p_1\ &\ 
	p_1 \actsfor (p_1 \lor p_2)\ &\ 
	\infer{p_1 \actsfor p_2\ \ \ \ p_2 \actsfor p_3}{p_1 \actsfor p_3}\ &\ 
	\infer{p_1 \actsfor p_3 \ \ \ \ p_2 \actsfor p_3}{(p_1 \lor p_2) \actsfor p_3}\ &\ 
	\infer{p_1 \actsfor p_2 \ \ \ \ p_1 \actsfor p_3}{p_1 \actsfor (p_2 \land p_3)}
	\end{array}
	\end{array}
	\]
\else
	\[
	\begin{array}{c}
	\begin{array}{c|c|c}
	(p_1 \land p_2) \actsfor p_1\ &\ 
	p_1 \actsfor (p_1 \lor p_2)\ &\ 
	\infer{p_1 \actsfor p_2\ \ \ \ p_2 \actsfor p_3}{p_1 \actsfor p_3}
	\end{array}\\
	\begin{array}{c|c}
	\infer{p_1 \actsfor p_3 \ \ \ \ p_2 \actsfor p_3}{(p_1 \lor p_2) \actsfor p_3}\ &\ 
	\infer{p_1 \actsfor p_2 \ \ \ \ p_1 \actsfor p_3}{p_1 \actsfor (p_2 \land p_3)}
	\end{array}
	\end{array}
	\]
\fi

\subsection{Labels}
Labels are sets of policies.

\subsubsection{$\join$ and $\meet$}
For availability or integrity labels:
\ifreport
	\[\begin{array}{r c l}
	I\p{\ell_1 \join \ell_2} &=& 
	  \left\{
	    \p{u_1\lor u_2} \integ \p{p_1 \lor p_2} \right| 
	\left.
	    u_1\integ p_1\in I\p{\ell_1} \land 
	    u_2\integ p_2\in I\p{\ell_2} \right\}
	\\
	A\p{\ell_1 \join \ell_2} &=& 
	  \left\{
	    \p{u_1\lor u_2} \avail \p{p_1 \lor p_2} \right| 
	\left.
	    u_1\avail p_1\in A\p{\ell_1} \land 
	    u_2\avail p_2\in A\p{\ell_2} \right\}
	\\
	\ell_1 \meet \ell_2 &=& \ell_1 \cup \ell_2 \end{array}
	\]
\else
	\begin{multline*}
	I\p{\ell_1 \join \ell_2} = 
	  \left\{
	    \p{u_1\lor u_2} \integ \p{p_1 \lor p_2} \right| \\
	\left.
	    u_1\integ p_1\in I\p{\ell_1} \land 
	    u_2\integ p_2\in I\p{\ell_2} \right\}
	\end{multline*}
	\vspace{-2em}
	\begin{multline*}
	A\p{\ell_1 \join \ell_2} = 
	  \left\{
	    \p{u_1\lor u_2} \avail \p{p_1 \lor p_2} \right| \\
	\left.
	    u_1\avail p_1\in A\p{\ell_1} \land 
	    u_2\avail p_2\in A\p{\ell_2} \right\}
	\end{multline*}
	\vspace{-3em}
	\begin{multline*}
	\ell_1 \meet \ell_2 = \ell_1 \cup \ell_2 \\
	\end{multline*}
\fi

\subsubsection{The lattice ordering}
$\less$ on labels is intuitively defined in Section~\ref{orderinglabelssection}.
Formally, it is defined in Figure ~\ref{fig:ordering}.
This definition is similar to Stephen Chong's~\cite{chong-thesis}.

\subsubsection{Equality}
Labels $\ell_1$ and $\ell_2$ are considered \textit{equal}, or at
least to lie in the same equivalence class, iff $\p{\ell_1 \less
\ell_2}\land\p{\ell_2\less\ell_1}$.

\begin{figure*}
\[
\frac{
\forall \cb{ p \in P } .\forall X\in\cb{A,I} . \p{
  \p{\wedge \cb{a \left|\p{p\actsfor o}\land\p{o{\xleftarrow{X}} a}\in \ell_1 \right.}} 
    \actsfor  
    \p{\wedge\cb{a \left|\p{p\actsfor o}\land\p{o{\xleftarrow{X}} a}\in \ell_2 \right.}}}
}{
\ell_1 \less \ell_2
}
\]
\caption{Ordering on labels.}\label{fig:ordering}
\ICS{This is really ugly. Anyone know how to make it prettier?}
\end{figure*}

\section{Heterogeneous consensus proofs}
\label{hetconsproofs}
\subsection{Agreement}
\label{agreementproof}
If two gurus decide, they can do so either in the same round (of fast 
consensus), or one can decide in fast consensus, and the other in 
{\underlyingconsensus}.

\subsubsection{Same Round:}
No two gurus decide different values in the same round, by 
$\p{\ref{eq:wrong-liar-change}}$, $\p{\ref{eq:threshold-order}}$, and 
$W \less A_{\textrm{attack}}\meet I_{attack}$ (from the definition of $W$). 
In particular, in order for a participant  to decide, a group of participants 
must send messages with a meet featuring a label $\less D$.
Therefore, the meet of labels of messages from participants who either lied to or did not
send the decided value to the first participant is $\notless C$. 
Given that $D \less C$, no message synthesized from 
participants who either lied to or did not send the decided value to the 
first participant is $\less D$.
Therefore, no participant can decide any value other than the one decided by the
first participant in the same round. 
Therefore, if two participants decide in the same round, they decide the same value.  

\subsubsection{Different Rounds:}
By $\p{\ref{eq:decide-change}}$, if one guru decides, then for each other
correct participant, there exists---among the participants from whom the guru has
received messages---some subset $J$ that is correct and whose messages
are received by the other participant.
Furthermore, $J$ is 
sufficient to change the vote of that other participant
to the value that has been decided. 

Therefore, all correct participants (who have not yet decided) enter 
{\underlyingconsensus} with the decided value as their starting value. 
Assuming the first deciding participant behaves at least as a participant in 
{\underlyingconsensus} (as is the case for the given decide procedure when 
{\underlyingconsensus} is fast consensus), then if {\underlyingconsensus} guarantees
unanimity, then all gurus will decide the same value as the first participant, 
and so all future decisions by gurus will agree with the first participant. 

This protocol does not permit the possibility of the same guru deciding twice 
in a round of fast consensus, and so agreement of {\underlyingconsensus}
combined with unanimity of {\underlyingconsensus} guarantees that any guru who
decides twice must decide the same value both times.

\subsection{Unanimity}
% note: I must prove this for the recursive case as well: no participant should decide other than v in {\underlyingconsensus}
\label{unanimityproof}
Given $\p{\ref{eq:threshold-order}}$, if all correct participants send messages 
of the same value, the meet of the labels of those messages is $\less D$.
By $\p{\ref{eq:decide-change}}$, this requires that even in the presence of 
attackers, all correct participants will receive a set of messages with the 
``correct'' value such that the meet of their labels $\less C$. 
All of the correct participants will therefore hold the same value when moving 
into {\underlyingconsensus}.

Therefore, if {\underlyingconsensus} has Unanimity, then so does Fast-consensus. 

If {\underlyingconsensus} is fast-consensus, then no correct participant can 
decide any value other than the correct value. 
Given that in each round, all correct participants will enter with the same
value, guaranteeing they do so in the next round, no correct participant will
ever broadcast any other value.
It is possible for a correct participant to receive all the messages from 
the correct participants first, and therefore decide on the correct value.

\subsection{Validity}
\label{validityproof}
The decision procedure of this protocol only allows a correct participant to 
decide on an element of the set of values from received messages. 
Because received messages must have been sent (network assumption), we have 
validity.

\subsection{Progress}
\label{progressproof}
From the perspective of any guru:

Given $\p{\ref{eq:message-avail-limit}}$ 
, $\p{\ref{eq:message-set-limit}}$, 
and $\p{\ref{eq:progress}}$,
any attacker with availability 
$\notless A_{\textrm{attack}}$
would be unable to violate the system availability  assumptions 
$\p{A_{\textrm{sys}}}$
of any set of participants such that the meet of the labels of the messages of
the remainder
$\notless A_{\textrm{sys}}$.

Therefore, any guru can always expect a set of messages such that the meet of 
the availability of their labels  $\less A_{\textrm{sys}}$, and can therefore always either 
decide or move on to {\underlyingconsensus}. 

\subsection{Termination}
\label{terminationproof}

If $p$ is a guru, then under the ``random network'' assumption, with
some non-zero probability $p$ will
in some round get messages from all correct processes, the meet of the labels of
which are $\less A_{\textrm{sys}}$, and so it will move on the next round.

There is likewise some non-zero probability that all correct participants
will receive messages in the same order as~$p$.

From requirement~\ref{eq:progress}, the availability of  the meet of the labels
of the set of
messages $p$ received must be enough to carry a set of participants into the
next round that will allow $p$ to make further progress.
(The combined availability of their messages to $p$ must be $\less A_{\textrm{sys}}$.)

From the structure of the protocol, and requirements
\ref{eq:wrong-liar-change} and \ref{eq:decide-change}, 
no two correct processes should be forced
to select different values after having received identical sets of
messages.

Therefore, if \textit{selection-function} has a non-zero probability of selecting
each item in the input set, there is a non-zero probability that
a round exists in which $p$ progresses, as do a set of other
correct participants who have sufficient availability for $p$ to continue
to progress, and all of them send messages of identical value.

By requirement~\ref{eq:threshold-order}, this is sufficient for $p$ to decide 
that value,
provided $p$ receives all of those messages first.
% (at least, under
% some interpretation of the pseudocode written, you could also read
% it such that you would always decide that value. It depends how you
% read "upon receipt of a message in . . .").
This will occur with some non-zero probability.
Therefore, in any pair of consecutive rounds, there is a non-zero
probability a guru will decide.
Therefore, each guru, with probability~1, eventually decides.

\newpage
\bibliography{bibtex/pm-master}

\begin{thebibliography}{40}
\providecommand{\natexlab}[1]{#1}
\providecommand{\url}[1]{\texttt{#1}}
\expandafter\ifx\csname urlstyle\endcsname\relax
  \providecommand{\doi}[1]{doi: #1}\else
  \providecommand{\doi}{doi: \begingroup \urlstyle{rm}\Url}\fi

\bibitem[Abraham et~al.(2006)Abraham, Dolev, Gonen, and Halpern]{abraham2006}
I.~Abraham, D.~Dolev, R.~Gonen, and J.~Halpern.
\newblock Distributed computing meets game theory: Robust mechanisms for
  rational secret sharing and multiparty computation.
\newblock In \emph{Proceedings of the Twenty-fifth Annual ACM Symposium on
  Principles of Distributed Computing}, PODC '06, pages 53--62, New York, NY,
  USA, 2006. ACM.

\bibitem[Afek et~al.(1994)Afek, Attiya, Fekete, Fischer, Lynch, Mansour, Wang,
  and Zuck]{Afek1994}
Y.~Afek, H.~Attiya, A.~Fekete, M.~Fischer, N.~Lynch, Y.~Mansour, D.-W. Wang,
  and L.~Zuck.
\newblock Reliable communication over unreliable channels.
\newblock \emph{J. ACM}, 41\penalty0 (6):\penalty0 1267--1297, Nov. 1994.
\newblock ISSN 0004-5411.
\newblock \doi{10.1145/195613.195651}.
\newblock URL \url{http://doi.acm.org/10.1145/195613.195651}.

\bibitem[Aiyer et~al.(2005)Aiyer, Alvisi, Clement, Dahlin, Martin, and
  Porth]{bar-sosp}
A.~S. Aiyer, L.~Alvisi, A.~Clement, M.~Dahlin, J.-P. Martin, and C.~Porth.
\newblock {BAR} fault tolerance for cooperative services.
\newblock In \emph{Proceedings of the Twentieth ACM Symposium on Operating
  Systems Principles}, SOSP '05, pages 45--58, New York, NY, USA, 2005. ACM.
\newblock ISBN 1-59593-079-5.
\newblock \doi{10.1145/1095810.1095816}.
\newblock URL \url{http://doi.acm.org/10.1145/1095810.1095816}.

\bibitem[Biba(1977)]{integrity}
K.~J. Biba.
\newblock Integrity considerations for secure computer systems.
\newblock Technical Report ESD-TR-76-372, USAF Electronic Systems Division,
  Bedford, MA, Apr. 1977.
\newblock (Also available through National Technical Information Service,
  Springfield Va., NTIS AD-A039324.).

\bibitem[Boyer and Moore(1984)]{boyer1984}
R.~S. Boyer and J.~S. Moore.
\newblock A mechanical proof of the unsolvability of the halting problem.
\newblock \emph{J. ACM}, 31\penalty0 (3):\penalty0 441--458, June 1984.
\newblock ISSN 0004-5411.
\newblock \doi{10.1145/828.1882}.
\newblock URL \url{http://doi.acm.org/10.1145/828.1882}.

\bibitem[Bracha and Toueg(1985)]{Bracha85}
G.~Bracha and S.~Toueg.
\newblock Asynchronous consensus and broadcast protocols.
\newblock \emph{Journal of the ACM}, 32\penalty0 (4):\penalty0 824--240, 1985.

\bibitem[Brasileiro et~al.(2001)Brasileiro, Greve, Mostefaoui, and
  Raynal]{Brasileiro2001}
F.~Brasileiro, F.~Greve, A.~Mostefaoui, and M.~Raynal.
\newblock Consensus in one communication step.
\newblock In V.~Malyshkin, editor, \emph{Parallel Computing Technologies},
  volume 2127 of \emph{Lecture Notes in Computer Science}, pages 42--50.
  Springer Berlin Heidelberg, 2001.
\newblock ISBN 978-3-540-42522-9.
\newblock \doi{10.1007/3-540-44743-1_4}.
\newblock URL \url{http://dx.doi.org/10.1007/3-540-44743-1_4}.

\bibitem[Castro and Liskov(1999)]{osdi99}
M.~Castro and B.~Liskov.
\newblock Practical {B}yzantine fault tolerance.
\newblock In \emph{Proc.~3rd Symposium on Operating Systems Design and
  Implementation}, New Orleans, LA, Feb. 1999.

\bibitem[Chong(2008)]{chong-thesis}
S.~Chong.
\newblock \emph{Expressive and Enforceable Information Security Policies}.
\newblock PhD thesis, Cornell University, Aug. 2008.

\bibitem[Chong and Myers(2006)]{cm06}
S.~Chong and A.~C. Myers.
\newblock Decentralized robustness.
\newblock In \emph{Proc.~19th {IEEE} Computer Security Foundations Workshop},
  pages 242--253, July 2006.

\bibitem[Foley(1991)]{foley1}
S.~N. Foley.
\newblock A taxonomy for information flow policies and models.
\newblock In \emph{Proc.~IEEE Symp. on Security and Privacy}, pages 98--108,
  1991.

\bibitem[Garay and Perry(1992)]{Garay1992}
J.~Garay and K.~Perry.
\newblock A continuum of failure models for distributed computing.
\newblock In A.~Segall and S.~Zaks, editors, \emph{Distributed Algorithms},
  volume 647 of \emph{Lecture Notes in Computer Science}, pages 153--165.
  Springer Berlin Heidelberg, 1992.
\newblock ISBN 978-3-540-56188-0.
\newblock \doi{10.1007/3-540-56188-9_11}.
\newblock URL \url{http://dx.doi.org/10.1007/3-540-56188-9_11}.

\bibitem[Ho et~al.(2007)Ho, Dolev, and Renesse]{Ho2007}
C.~Ho, D.~Dolev, and R.~Renesse.
\newblock Making distributed applications robust.
\newblock In E.~Tovar, P.~Tsigas, and H.~Fouchal, editors, \emph{Principles of
  Distributed Systems}, volume 4878 of \emph{Lecture Notes in Computer
  Science}, pages 232--246. Springer Berlin Heidelberg, 2007.
\newblock ISBN 978-3-540-77095-4.
\newblock \doi{10.1007/978-3-540-77096-1_17}.
\newblock URL \url{http://dx.doi.org/10.1007/978-3-540-77096-1_17}.

\bibitem[Ho et~al.(2008)Ho, Van~Renesse, Bickford, and Dolev]{Ho2008}
C.~Ho, R.~Van~Renesse, M.~Bickford, and D.~Dolev.
\newblock Nysiad: Practical protocol transformation to tolerate {B}yzantine
  failures.
\newblock In \emph{Proc.~5th {USENIX} Symp.~on Networked Systems Design and
  Implementation ({NSDI})}, volume~8, pages 175--188, 2008.

\bibitem[Jaffe et~al.(2012)Jaffe, Moscibroda, and Sen]{jaffe2012}
A.~Jaffe, T.~Moscibroda, and S.~Sen.
\newblock On the price of equivocation in {B}yzantine agreement.
\newblock In \emph{Proceedings of the 2012 ACM Symposium on Principles of
  Distributed Computing}, PODC '12, pages 309--318, New York, NY, USA, 2012.
  ACM.
\newblock ISBN 978-1-4503-1450-3.
\newblock \doi{10.1145/2332432.2332491}.
\newblock URL \url{http://doi.acm.org/10.1145/2332432.2332491}.

\bibitem[Junqueira and Marzullo(2003)]{survivor-sets}
F.~Junqueira and K.~Marzullo.
\newblock Designing algorithms for dependent process failures.
\newblock In \emph{Proceedings of the Workshop on Future Directions in
  Distributed Computing}, pages 24--28, 2003.

\bibitem[Junqueira and Marzullo(2005)]{Junqueira2005}
F.~Junqueira and K.~Marzullo.
\newblock Replication predicates for dependent-failure algorithms.
\newblock In J.~Cunha and P.~Medeiros, editors, \emph{Euro-Par 2005 Parallel
  Processing}, volume 3648 of \emph{Lecture Notes in Computer Science}, pages
  617--632. Springer Berlin Heidelberg, 2005.
\newblock ISBN 978-3-540-28700-1.
\newblock \doi{10.1007/11549468_69}.
\newblock URL \url{http://dx.doi.org/10.1007/11549468_69}.

\bibitem[Lamport(1998)]{paxos}
L.~Lamport.
\newblock The {P}art-time {P}arliament.
\newblock \emph{ACM Trans. Comput. Syst.}, 16\penalty0 (2):\penalty0 133--169,
  May 1998.
\newblock ISSN 0734-2071.
\newblock \doi{10.1145/279227.279229}.
\newblock URL \url{http://doi.acm.org/10.1145/279227.279229}.

\bibitem[Lamport(2006)]{fastpaxos}
L.~Lamport.
\newblock Fast {Paxos}.
\newblock \emph{Distributed Computing}, 19\penalty0 (2):\penalty0 79--103,
  October 2006.
\newblock URL
  \url{http://research.microsoft.com/apps/pubs/default.aspx?id=64624}.

\bibitem[Lamport et~al.(1982)Lamport, Shostak, and Pease]{byzantinegenerals}
L.~Lamport, R.~Shostak, and M.~Pease.
\newblock The {B}yzantine generals problem.
\newblock \emph{ACM Trans. Program. Lang. Syst.}, 4\penalty0 (3):\penalty0
  382--401, July 1982.
\newblock ISSN 0164-0925.
\newblock \doi{10.1145/357172.357176}.
\newblock URL \url{http://doi.acm.org/10.1145/357172.357176}.

\bibitem[Lampson et~al.(1991)Lampson, Abadi, Burrows, and Wobber]{labw91}
B.~Lampson, M.~Abadi, M.~Burrows, and E.~Wobber.
\newblock Authentication in distributed systems: Theory and practice.
\newblock In \emph{Proc.~13th {ACM} Symp.~on Operating System Principles
  (SOSP)}, pages 165--182, Oct. 1991.
\newblock {\em Operating System Review}, 253(5).

\bibitem[Malkhi and Reiter(1997)]{Malkhi97a}
D.~Malkhi and M.~Reiter.
\newblock {B}yzantine quorum systems.
\newblock In \emph{Proc.~29th ACM Symposium on Theory of Computing}, pages
  569--578, El Paso, Texas, May 1997.

\bibitem[Meyer and Pradhan(1991)]{Meyer1991}
F.~Meyer and D.~Pradhan.
\newblock Consensus with dual failure modes.
\newblock \emph{Parallel and Distributed Systems, IEEE Transactions on},
  2\penalty0 (2):\penalty0 214--222, Apr 1991.
\newblock ISSN 1045-9219.
\newblock \doi{10.1109/71.89066}.

\bibitem[Myers and Liskov(2000)]{ml-tosem}
A.~C. Myers and B.~Liskov.
\newblock Protecting privacy using the decentralized label model.
\newblock \emph{ACM Transactions on Software Engineering and Methodology},
  9\penalty0 (4):\penalty0 410--442, Oct. 2000.
\newblock URL \url{http://www.cs.cornell.edu/andru/papers/iflow-tosem.pdf}.

\bibitem[Myers et~al.(2006)Myers, Zheng, Zdancewic, Chong, and Nystrom]{jif}
A.~C. Myers, L.~Zheng, S.~Zdancewic, S.~Chong, and N.~Nystrom.
\newblock {Jif 3.0}: {J}ava information flow.
\newblock Software release, \url{http://www.cs.cornell.edu/jif}, July 2006.

\bibitem[Palsberg and \O{}rb\ae{}k(1995)]{po95}
J.~Palsberg and P.~\O{}rb\ae{}k.
\newblock Trust in the $\lambda$-calculus.
\newblock In \emph{Proc.~2nd International Symposium on Static Analysis},
  number 983 in Lecture Notes in Computer Science, pages 314--329. Springer,
  Sept. 1995.

\bibitem[Postel(1981)]{TCP}
J.~Postel.
\newblock {DoD} standard transmission control protocol.
\newblock DARPA-Internet RFC-793, Sept. 1981.

\bibitem[Sabelfeld and Myers(2003)]{sm-jsac}
A.~Sabelfeld and A.~C. Myers.
\newblock Language-based information-flow security.
\newblock \emph{IEEE Journal on Selected Areas in Communications}, 21\penalty0
  (1):\penalty0 5--19, Jan. 2003.
\newblock URL \url{http://www.cs.cornell.edu/andru/papers/jsac/sm-jsac03.pdf}.

\bibitem[Schlichting and Schneider(1983)]{Schneider83a}
R.~D. Schlichting and F.~B. Schneider.
\newblock Fail-stop processors: An approach to designing fault-tolerant
  computing systems.
\newblock \emph{ACM Transactions on Computer Systems}, 1\penalty0 (3):\penalty0
  222--238, 1983.

\bibitem[Schneider(1990)]{Schneider90}
F.~B. Schneider.
\newblock Implementing fault-tolerant services using the state machine
  approach: a tutorial.
\newblock \emph{ACM Computing Surveys}, 22\penalty0 (4):\penalty0 299--319,
  Dec. 1990.

\bibitem[Siu et~al.(1998)Siu, Chin, and Yang]{Siu1998}
H.-S. Siu, Y.-H. Chin, and W.-P. Yang.
\newblock Byzantine agreement in the presence of mixed faults on processors and
  links.
\newblock \emph{Parallel and Distributed Systems, IEEE Transactions on},
  9\penalty0 (4):\penalty0 335--345, Apr 1998.
\newblock ISSN 1045-9219.
\newblock \doi{10.1109/71.667895}.

\bibitem[Song and Renesse(2008)]{Song2008}
Y.~J. Song and R.~Renesse.
\newblock Bosco: One-step {B}yzantine asynchronous consensus.
\newblock In \emph{Proc.~22nd Int'l Symp. on Distributed Computing}, DISC '08,
  pages 438--450, Berlin, Heidelberg, 2008. Springer-Verlag.
\newblock ISBN 978-3-540-87778-3.
\newblock \doi{10.1007/978-3-540-87779-0_30}.
\newblock URL \url{http://dx.doi.org/10.1007/978-3-540-87779-0_30}.

\bibitem[Steiner(1993)]{Steiner1993}
P.~Steiner.
\newblock On the {I}nternet, nobody knows you're a dog.
\newblock \emph{The New Yorker}, 69\penalty0 (20):\penalty0 61, 1993.
\newblock URL \url{http://www.unc.edu/depts/jomc/academics/dri/idog.html}.

\bibitem[Walker et~al.(2006)Walker, Mackey, Ligatti, Reis, and
  August]{lambda-zap}
D.~Walker, L.~Mackey, J.~Ligatti, G.~Reis, and D.~August.
\newblock Static typing for a faulty lambda calculus.
\newblock In \emph{ACM SIGPLAN International Conference on Functional
  Programming}, Sept. 2006.

\bibitem[Zdancewic et~al.(2002)Zdancewic, Zheng, Nystrom, and Myers]{zznm02}
S.~Zdancewic, L.~Zheng, N.~Nystrom, and A.~C. Myers.
\newblock Secure program partitioning.
\newblock \emph{ACM Transactions on Computer Systems}, 20\penalty0
  (3):\penalty0 283--328, Aug. 2002.
\newblock URL \url{http://www.cs.cornell.edu/andru/papers/sosp01/spp-tr.pdf}.

\bibitem[Zhang and Myers(2014)]{zm14}
D.~Zhang and A.~C. Myers.
\newblock Toward general diagnosis of static errors.
\newblock In \emph{Proc.~41st {ACM} Symposium on Principles of Programming
  Languages (POPL)}, pages 569--581, Jan. 2014.
\newblock URL \url{http://www.cs.cornell.edu/andru/papers/diagnostic}.

\bibitem[Zheng(2007)]{zlt-thesis}
L.~Zheng.
\newblock \emph{Making distributed computation secure by construction}.
\newblock PhD thesis, Cornell University, Ithaca, New York, USA, Jan. 2007.

\bibitem[Zheng and Myers(2005)]{zm05}
L.~Zheng and A.~C. Myers.
\newblock End-to-end availability policies and noninterference.
\newblock In \emph{Proc.~18th {IEEE} Computer Security Foundations Workshop},
  pages 272--286, June 2005.

\bibitem[Zheng and Myers(2014)]{Zheng2014}
L.~Zheng and A.~C. Myers.
\newblock A language-based approach to secure quorum replication.
\newblock In \emph{Proceedings of the Ninth Workshop on Programming Languages
  and Analysis for Security}, PLAS'14, pages 27:27--27:39, New York, NY, USA,
  2014. ACM.
\newblock ISBN 978-1-4503-2862-3.
\newblock \doi{10.1145/2637113.2637117}.
\newblock URL \url{http://doi.acm.org/10.1145/2637113.2637117}.

\bibitem[Zheng et~al.(2003)Zheng, Chong, Myers, and Zdancewic]{zcmz03}
L.~Zheng, S.~Chong, A.~C. Myers, and S.~Zdancewic.
\newblock Using replication and partitioning to build secure distributed
  systems.
\newblock In \emph{Proc.~IEEE Symp. on Security and Privacy}, pages 236--250,
  May 2003.
\newblock URL \url{http://www.cs.cornell.edu/andru/papers/sp03.pdf}.

\end{thebibliography}

\end{document}